\documentclass[acmsmall]{acmart}

\acmJournal{PACMPL}

\setcopyright{none}

\usepackage{booktabs}   
\usepackage{subcaption} 
\usepackage{color}
\usepackage{microtype}
\usepackage{amsmath}
\usepackage{graphicx}
\usepackage{amsthm}
\usepackage{stmaryrd}
\usepackage[all]{xy}
\usepackage{multirow}
\usepackage{paralist}
\usepackage{hhline}
\usepackage{bm}
\usepackage{listings,lstcoq}
\definecolor{ltblue}{rgb}{0,0.4,0.4}
\definecolor{dkblue}{rgb}{0,0.1,0.6}
\definecolor{dkgreen}{rgb}{0,0.35,0}
\definecolor{dkviolet}{rgb}{0.3,0,0.5}
\definecolor{dkred}{rgb}{0.5,0,0}
\usepackage{pmboxdraw}
\usepackage{pifont}
\newcommand{\cmark}{\CIRCLE}%
\newcommand{\xmark}{\Circle}%
\renewcommand{\matrix}[1]{\begin{bmatrix}#1\end{bmatrix}}
\usepackage{wasysym}
\usepackage{extarrows}
\usepackage{tikz}
\usepackage{wrapfig}

\usepackage{amsmath}
\usepackage{amsthm}
\usepackage{xspace}

\newtheorem{theorem}{Theorem}[section]

\newtheorem{lemma}{Lemma}[section]
\newtheorem{remark}{Remark}[section]

\newtheorem{example}{Example}[section]

\newtheorem{postulate}{Postulate}

\newtheorem{defn}{Definition}[section]

\newcommand {\bC } {{\mathbb{C}}}

\newcommand {\bZ } {{\mathbb{Z}}}
\newcommand {\bF } {{\mathbb{F}}}

\newcommand {\bu } {{\mathbf{u}}}
\newcommand {\bv } {{\mathbf{v}}}
\newcommand {\bw } {{\mathbf{w}}}

\renewcommand{\bra}[1]{\langle #1|}
\renewcommand{\ket}[1]{|#1\rangle}

\newcommand {\cD } {{\mathcal{D}}}
\newcommand {\cE } {{\mathcal{E}}}

\newcommand {\cH } {{\mathcal{H}}}
\newcommand {\cI } {{\mathcal{I}}}

\newcommand {\cL } {{\mathcal{L}}}
\newcommand {\cM } {{\mathcal{M}}}

\newcommand {\cO } {{\mathcal{O}}}
\newcommand {\cP } {{\mathcal{P}}}

\newcommand {\cU } {{\mathcal{U}}}

\newcommand {\cSO} {{\mathcal{SO}}}
\newcommand {\cQO} {{\mathcal{QO}}}
\newcommand {\cQC} {{\mathcal{QC}}}
\newcommand {\cCP} {{\mathcal{CP}}}

\newcommand{\cg}[1]{{{\color{purple}\left\{{\color{black}#1}\right\}}}}
\newcommand{\cgrule}[5]{{{\color{purple}\models}_{#1}^{#2}\cg{#3}{#4}\cg{#5}}}

\newcommand {\codoms }[1] {{\mathsf{codom}}}

\newcommand {\tr } {{\mathrm{tr}}}

\newcommand {\sem}[1] {\llbracket#1\rrbracket}
\newcommand {\lsem}[1] {\llbracket#1\rrbracket_l}
\newcommand {\coqm}[1] {\text{\small\ttfamily{#1}}}
\newcommand{\mx}{{{\tt x}}}
\newcommand{\my}{{{\tt y}}}
\newcommand{\mt}{{{\tt t}}}
\newcommand{\ms}{{{\tt s}}}
\newcommand{\mT}{{{\tt T}}}
\newcommand{\mi}{{{\tt i}}}
\newcommand{\mj}{{{\tt j}}}

\newcommand{\mn}{{{\tt n}}}
\renewcommand{\mp}{{{\tt p}}}
\newcommand{\mq}{{{\tt q}}}
\newcommand{\mr}{{{\tt r}}}

\newcommand {\spans } {{\mathrm{span}}}

\def\>{\ensuremath{\rangle}}
\def\<{\ensuremath{\langle}}
\def\ra{\ensuremath{\rightarrow}}

\DeclareTextCommand{\tus}{OT1}{%
  \leavevmode \kern.06em\vbox{\hrule width.6em}}

\renewcommand{\tt}[1]{\texttt{\frenchspacing{#1}}}

\newcommand{\kif}{{\mathbf{if}}}
\newcommand{\kwhile}{{\mathbf{while}}}
\newcommand{\kfi}{{\mathbf{fi}}}

\newcommand{\icond}[4]{{\mathbf{if}\ {#2}[{#3}] = {#1}\rightarrow{#4}_{#1}\ \mathbf{fi}}}
\newcommand{\iwhile}[4]{{\mathbf{while}\ {#1}[{#2}] = {#3}\ \mathbf{do}\ {#4}\ \mathbf{od}}}

\newcommand{\aabort}{{\mathbf{abort}}}
\newcommand{\askip}{{\mathbf{skip}}}
\newcommand{\pset}[1]{{\text{\textrm{set}}({#1})}}
\newcommand{\ainit}[1]{{\mathbf{init}\ {#1}}}
\newcommand{\aapply}[1]{{\mathbf{apply}\ {#1}}}
\newcommand{\acond}[3]{{\mathbf{if}\ (\square{#1}\cdot{#2} = {#1}\rightarrow{#3}_{#1})\ \mathbf{fi}}}
\newcommand{\awhile}[3]{{\mathbf{while}\ {#1} = {#2}\ \mathbf{do}\ {#3}\ \mathbf{od}}}
\newcommand{\akwhile}[4]{{\mathbf{while}^{(#4)}\ {#1} = {#2}\ \mathbf{do}\ {#3}\ \mathbf{od}}}
\newcommand{\afor}[3]{{\mathbf{for}\ {#1}{#2}\ \mathbf{do}\ {#3}_{#1}}}

\newcommand{\cl}{{{cl}}}
\renewcommand{\imath}{\mathbf{i}}

\begin{document}

\title{CoqQ : Foundational Verification of Quantum Programs}


\author{Li Zhou}
\affiliation{
  \institution{Max Planck Institute for Security and Privacy (MPI-SP)}           
}
\email{li.zhou@mpi-sp.org}
\author{Gilles Barthe}
\affiliation{
  \institution{Max Planck Institute for Security and Privacy (MPI-SP)}            
}
\affiliation{
  \institution{IMDEA Software Institute}            
}
\email{gjbarthe@gmail.com}
\author{Pierre-Yves Strub}
\affiliation{
  \department{None}             
  \institution{None}           
}
\email{pierre-yves@strub.nu}
\author{Junyi Liu}
\affiliation{
  \institution{State Key Laboratory of Computer Science, Institute of Software, Chinese Academy of Sciences, China}           
}
\affiliation{
  \institution{University of Chinese Academy of Sciences, China}           
}
\email{liujy@ios.ac.cn}
\author{Mingsheng Ying}
\affiliation{
  \institution{State Key Laboratory of Computer Science, Institute of Software, Chinese Academy of Sciences, China}           
}
\affiliation{
  \institution{Tsinghua University, China}           
}
\email{yingms@ios.ac.cn}

\begin{abstract}
  CoqQ is a framework for reasoning about quantum programs in the Coq
  proof assistant. Its main components are: a deeply embedded quantum
  programming language, in which classic quantum algorithms are easily
  expressed, and an expressive program logic for proving properties of
  programs. CoqQ is foundational: the program logic is formally proved
  sound with respect to a denotational semantics based on state-of-art
  mathematical libraries (mathcomp and mathcomp analysis). CoqQ is
  also practical: assertions can use Dirac expressions, which eases
  concise specifications, and proofs can exploit local and parallel
  reasoning, which minimizes verification effort. We illustrate the
  applicability of CoqQ with many examples from the literature.
  
\end{abstract}



\maketitle

\section{Introduction}
Quantum programming languages hold the promise of making the computational power of quantum computers readily accessible to software developers. As such, they are a key element in large-scale efforts to deploy quantum computing.  Examples of industrially supported quantum programming languages include  IBM's Qiskit~\cite{Qiskit}, Google's Cirq~\cite{Cirq} and Microsoft's~Q\# \cite{Qsharp}. In spite of the benefits of quantum programming languages, writing correct quantum programs remains error-prone, not the least because quantum computation often contradicts human intuition.

(Deductive) program verification provides a principled approach for reasoning compositionally about programs. One fundamental strength of deductive program verification is its ability to reason about rich specifications, and thus to prove that programs are correct. Deductive program verification is traditionally based on program logics. Informally, these program logics feature a syntax-directed proof rules, which can be used (backwards) to decompose proof goals into simpler goals. Ultimately the simplest goals can be established in isolation using (e.g.,\, first-order predicate) logic. Program verification and program logics are used extensively by the software industry to validate large-scale classic developments. Similarly, there exist several program logics for reasoning about quantum programs (see Section~\ref{sec:related}). These logics obey the same principles as their counterparts for classic programs. Yet, they do not achieve similar usability and scalability.

Program logics for quantum programs face two main challenges in comparison to program logics for classic programs. First, quantum states and assertions have a richer structure than their classic counterparts. In particular, quantum states have the structure of a Hilbert space and quantum assertions are modelled as Hermitian operators~\cite{DP06}. As a consequence, proving entailment of assertions, as required for instance by the rule of consequence, often involves complex calculations. Second, the effect of atomic instructions on quantum states is inherently non-local, because these states may be entangled. As a consequence, proof rules for quantum programs do not have natural support for local reasoning, such as the classic frame rule in Hoare logic. This makes reasoning very cumbersome and error-prone.

A common means to ensure that proofs of program correctness adhere to the program logic is to provide mechanized support for the program logic. Mechanized support is beneficial in many ways: it can help book keeping proof obligations and discharging boilerplate obligations automatically.  However, when program logics are themselves very complex, there is the risk that the implementation of the program logic is itself flawed. One appealing approach to address these concerns is to develop verified program verifiers, i.e.\, program verifiers which are themselves proved correct w.r.t.\, the semantics of programs. A verified program verifier typically consists of several ingredients:
\begin{itemize}
\item an embedding of the programming language. One often opts for a deep embedding, in which the syntax of programs is modelled as an inductive type;

\item a semantics of the programming language.  The semantics is built from first principles and defines the behavior of programs;

\item a formalization of the program logic. Typically, each rule of the program logic is formalized as a lemma showing the soundness of the rule w.r.t.\, the semantics.
\end{itemize}
As a consequence, a program verified using the formalized program logic can be deemed correct from first principles. There exist many verified verifiers for classical programming languages, including the Verified Software Toolchain (VST)~\cite{vst} for C programs---which in addition integrates the CompCert verified compiler~\cite{compcert}---and Iris/RustBelt for Rust programs~\cite{iris,rustbelt}.

\paragraph*{Contributions}
In this paper, we present CoqQ, a verified program verifier for \textbf{qwhile}, a core imperative language with classic control flow and quantum data. As its name suggests, CoqQ is formalized in the Coq proof assistant, which has been used extensively to formalize mathematics and program semantics. The main components of CoqQ
are:

\begin{itemize}
\item a formalization of the \textbf{qwhile} programming language. The crux of our formalization is a denotational semantics built from first principles. Specifically, we extend the algebraic hierarchy of mathcomp~\cite{mathcomp}, an extensive library of formalized mathematics, with a formalization of Hilbert spaces.  Such a formalization
  of Hilbert spaces is essential for generality of the approach and for enabling local reasoning, as we discuss below. Note that our formalization is restricted to finite-dimensional Hilbert spaces; we inherit this limitation from mathcomp's formalization of vector spaces, which is also restricted to finite dimension.  However, this restriction is not a problem for verifying the existing quantum algorithms;

\item a formalization of labelled Dirac notation. Dirac notation, a.k.a.\, bra - ket notation, is used ubiquitously to model quantum states. One major benefit of Dirac notation is to support equational reasoning about quantum systems. However, it does not reflect (and hence does not exploit) the structure of quantum states. To overcome this limitation, quantum physicists, in particular those working on many-body quantum physics, commonly use labels to tag quantum expressions with the (sub)systems in which they live or operate. The benefit of the resulting labelled Dirac notation is that it avoids the need of writing the matrix representation of a large state, an observable or a Hamiltonian of a many-body system; instead, it allows to define the above objects as a linear combination of the tensor products of local quantities with labels (e.g. subscripts) to indicate the involved subsystems.

CoqQ provides a formalization of labelled Dirac notation; to our best knowledge, no such formalization has been developed before. The key element of the formalization is an interpretation of expressions written using labelled Dirac notation using abstract tensor products of Hilbert spaces. More specifically, given a finite set of symbols $L$ and the abstract Hilbert space $\cH_x$ for each $x\in L$ (the state Hilbert space of each quantum variable), we define the Hilbert space $\cH_S\triangleq\bigotimes_{x\in S}\cH_x$ corresponding to any subset $S\subseteq L$. We use this tensor product construction to interpret labelled Dirac notation. One technical issue is that a complex expression may be interpreted in a different Hilbert space than its immediate subexpressions. Our formalization addresses this issue by leveraging canonical structures and big operators, two classic tools already used extensively by the mathcomp libraries.

We leverage our formalization to validate identities that are commonly used when calculating using labelled Dirac notation, e.g.\, the following commutativity property:
  $$|\phi\>_{S_1}|\psi\>_{S_2} = |\psi\>_{S_2}|\phi\>_{S_1} \qquad \mbox{if} \qquad S_1\cap S_2 = \emptyset$$

\item a verified Hoare logic. Statements of the logic are of the form $\{A_{S_1}\}C\{B_{S_2}\}$, where $C$ is a quantum program, $A$ and $B$ are Hermitian operators over the subspaces defined by $S_1$ and $S_2$ respectively. Following a standard practice in verified program verifiers, every proof rule is formalized as a lemma that is stated relative to the denotational semantics of its corresponding construct. One specificity of our program logic is that it allows to express program correctness as state transformation. Specifically, our formalization supports judgments of the form (we use purple instead of introducing new notation):
 $$\cgrule{}{}{|u\>_{S_u}}{C}{|v\>_{S_v}}$$
 stating that $C$ transforms the input state $|u\>_{S_u}$ to the output state $|v\>_{S_v}$, assuming that $\||u\>_{S_u}\| = \||v\>_{S_v}\| = 1$. This kind of judgment provides a \emph{human-readable}  statement that looks akin to statements found in textbook presentations of quantum algorithms.
\end{itemize}
In order to illustrate the benefits of CoqQ, we conduct a representative set of case studies, including HHL algorithm for solving linear equations \cite{HHL09}, Grover's search algorithm \cite{Gro96}, quantum phase estimation (QPE) and the hidden subgroup problem (HSP) algorithm \cite{Kit95,Lom04}, together with the circuit implementation of quantum Fourier transformation (QFT) and Bravyi-Gosset-Konig’s algorithm for hidden linear function (HLF) problem \cite{BGK18}. Several of these examples have been verified before; one notable exception is the HSP algorithm, which requires the full generality of \textbf{qwhile} and cannot be expressed (let alone be verified) easily in other formalisms.

The development of CoqQ can be found at \url{https://github.com/coq-quantum/CoqQ}.

\paragraph*{Related work} There is a large body of work in the design, implementation, and verification of quantum programs. However, CoqQ is to our best knowledge the first formally verified program verifier for a high-level quantum programming language that operates over a general notion of state. We elaborate  below, by contrasting CoqQ with other tools that support verification of quantum programs within proof assistants---a more extensive and detailed comparison is found in Section~\ref{sec:related}. The comparison is summarized pictorially in Table~\ref{tab:related-work}.

Our work is most closely related to formalizations of \textbf{qwhile}. There are two such formalizations. The first one is QHLProver~\cite{liu2019formal}, which is used for proving correctness of quantum programs based on quantum Hoare logic. The formalization is based on a representation of quantum states as matrices (using the Jordan Normal Form library from Isabelle~\cite{thiemann2016formalizing}), rather than on a general theory of Hilbert spaces. The second one is qrhl~\cite{Unr19,Unr20}, which is used for relational verification of quantum programs. The formalization of qrhl is not (yet) foundational. Rather, the proof rules of the logic are modelled as axioms. 

However, the currently prevailing line of work uses proof assistants to reason about circuit-based quantum programs that operate over concrete representations of quantum states. There are several efforts in this direction, including QWIRE~\cite{paykin2017qwire,RPZ18}, SQIR~\cite{HRH21,hietala2020proving} and IMD (Isabelle Marries Dirac)~\cite{bordg2021certified}. These efforts are rather different from ours, and elide several of the key challenges addressed by CoqQ.

\begin{table}
\begin{tabular}{|c|c|c|c|c|c|c|} 
\hline
 Tool & Programming model & State & Ax.Sem. & Found. & Proof Assistant & Libraries \\
  \hline
  \hline
QWIRE     & Circuit & \xmark & \xmark & \cmark & Coq & Std. Lib. \\
\hline
SQIR       &  Circuit & \xmark & \xmark & \cmark & Coq  & Std. Lib. \\
\hline
QHLProver   & High-level & \xmark & \cmark & \cmark & Isabelle & JNF, DL \\
  \hline
  qrhl-tool & High-level & \cmark  & \cmark$\mbox{}^{\ddagger}$ & \xmark & Isabelle & JNF, CBO \\
\hline
IMD   & Mathematics & \xmark & \xmark & \cmark & Isabelle & JNF \\
\hline

\hline
\hline
CoqQ   & High-level & \cmark & \cmark & \cmark & Coq & MathComp \\
\hline
\end{tabular}
\caption{Comparison with other mechanizations of quantum programming languages. The column (State) indicates if the formalization supports a general notion of state. The column (Ax. Sem.) indicates if the formalization includes a program logic. The column (Found.) means the formalization is built from first principles. $\mbox{}^\ddagger$ indicates that qrhl-tool supports relational verification.
  Libraries are discussed in Section~\ref{sec:related}.}
\label{tab:related-work}
\end{table}

As usual, the tools cannot be ordered by comparing the number of $\cmark$ in the table. In particular, there are evident trade-offs between high-level programming languages and
circuits. Approaches based on the former can typically ease the specification and verification of high-level algorithms, whereas approaches based on the latter remain closer to implementations, and blend more nicely with verified compilers. We discuss these trade-offs further in Section~\ref{sec:related}.

\section{Motivating Example: Hidden Subgroup Problem}
\label{sec-motivation}
Let $G$ be a finite group and $Y$ be a finite set. The Hidden Subgroup Problem (HSP) \cite{Kit95} is to compute, given oracle access to a function $f:G\rightarrow Y$, its hidden subgroup structure. More precisely, assuming that there exists a subgroup $H$ of $G$ such that for every $g_1, g_2 \in G$, $f(g_1) = f(g_2)$ if and only if $g_2 -  g_1 \in H$; the Hidden Subgroup Problem (HSP) is to compute a subset $Z\subseteq G$ such that its generated subgroup $\langle Z \rangle$ satisfies $\langle Z \rangle=H$. The problem arises in many settings, including integer factorization, discrete logarithm,
and graph theory.

\begin{wrapfigure}{r}{5cm}
$\begin{array}{l}
\coqm{\textbf{for} $0 \leq i < k$ \textbf{do} x$_i$ := $|0\>$;}\\
\coqm{y := $|$y$_0\>$ ;} \\
\coqm{\textbf{for} $0 \leq i < k$ \textbf{do} x$_i$ := QFT[x$_i$];} \\ 
\coqm{[$\overline{\mx}$, y] := $U[f]$[$\overline{\mx}$, y];} \\
\coqm{\textbf{for} $0 \leq i < k$ \textbf{do} x$_i$ := QFT[x$_i$].}
\end{array}$
\caption{HSP algorithm}
\label{fig:hsp:code}
\end{wrapfigure}

The existence of polynomial-time quantum algorithms for solving HSP over arbitrary finite groups remains an open problem. However, there are polynomial-time quantum algorithms for solving HSP over finite abelian groups. The core of these algorithms is a quantum procedure that samples uniformly over the orthogonal subgroup $H^\perp$ of $H$. The subgroup is defined as
$H^\perp = \{ g \in G\ |\ \forall h\in H.~\chi_g(h)= 1\}$,
where $\chi_g:G\rightarrow \mathbb{C}^*$ is a so-called character (defined formally in Section~\ref{sec:hsp}; definition is irrelevant for this overview)---recall that $\mathbb{C}^*$ is the set of non-zero complex numbers. Given a procedure for sampling uniformly from $H^\perp$, one can solve the Hidden Subgroup Problem by generating sufficiently many samples, and by applying a classic algorithm to convert a generating set for $H^\perp$ into a generating set for $H$. We omit the details here and refer the reader to~\cite{NC00,Lom04} for further details.

The code $HSP$ of the sampling algorithm appears in Figure~\ref{fig:hsp:code}.  The code operates over two quantum registers: the register $\overline{\mx}$ ranges over the cyclic group decomposition $\mathbb{Z}_{p_0} \times \ldots \times \mathbb{Z}_{p_{k-1}}$ of $G$ and the register $\my$ ranges over the finite set $Y$. We let $\mx_i$ denote the $i^{\mathrm{th}}$ projection of $\overline{\mx}$. We note that this ability to declare rich types for variables simplifies the the writing of the algorithm---and ultimately eases verification. We briefly comment on the code:

\begin{itemize}
\item the first two lines initialize the registers $\overline{\mx}$ and $\my$ to default values; 
\item the second \textsf{\textbf{for}} loop applies QFT (Quantum Fourier Transform) on every $\mx_i$; 
\item the next line applies the unitary transformation $U[f]$ that provides quantum access to $f$;
\item the third \textsf{\textbf{for}} loop applies QFT on every $\mx_i$.
\end{itemize}
Correctness of $HSP$ is states informally as follows: $HSP$ samples uniformly from the so-called orthogonal subgroup $H^\perp $ of $H$---the formal definition of $H^\perp $ is deferred to Section~\ref{sec:hsp}. However, assertions in quantum Hoare logic are Hermitian operators that model observables of the quantum system. In order to capture the correctness of $HSP$ formally, we consider the assertions (parametrized by $g\in G$) $\left\{|g\>_{\overline{\mx}}\<g|\right\}$, which represents the probability of observing $g$ when measuring $\overline{\mx}$ on the output state.  Correctness is captured by the following statements (for simplicity, we omit proof modes that will be discussed later):
\begin{align}
    & \forall\, g\in H^\perp,\ \models \left\{\frac{1}{|H^\perp|}\right\}\ HSP\ \left\{|g\>_{\overline{\mx}}\<g|\right\}\label{eqn-triple-HSP-in}\\
    & \forall\, g\notin H^\perp,\ \models \left\{0\right\}\ HSP\ \left\{|g\>_{\overline{\mx}}\<g|\right\}\label{eqn-triple-HSP-notin}.
\end{align}
Taken together, these statements entail that $HSP$ outputs an element of $H^\perp$ uniformly at random. The derivation of these triples in quantum Hoare logic proceeds structurally, following the order of execution:
\begin{itemize}
\item we use local and parallel reasoning to reason about initialization of the registers;
\item we use local and parallel reasoning, and some auxiliary proof about QFT, to reason about application of QFT. Moreover, we use basic facts from group theory to
transform the assertion into an equivalent one;
\item we use the unitary rule to reason about the oracle call. Moreover, we use basic facts from group theory to transform the assertion into an equivalent one;
\item we use local and parallel reasoning, and some auxiliary proof about QFT, to reason about application of QFT. Moreover, we use basic facts from group theory to
  transform the assertion into an equivalent one;
\item last, we apply a new rule, called  (R.Inner), to establish the desired Hoare triple. At a high-level, the rule helps combining backwards and forward reasoning, and helps to write formal proofs that follow informal proofs closely.
\end{itemize}
Note that each instruction combines an application of the proof rule, proof obligations to justify local/parallel reasoning, and domain-specific reasoning (here group theory) to recast intermediate assertions into a suitable form. This makes the proof intricate, and a good example for mechanization in a verified program verifier. We provide additional details in Section~\ref{sec:hsp}.

\section{Preliminaries}\label{sec:prelim}

Quantum computing is built upon quantum mechanics (the finite-dimensional instance of) which might be characterized by linear algebra. We will briefly review some basic definitions of linear algebra and then a quick introduction how quantum mechanics are formalized using concepts in linear algebra.

\subsection{Abstract linear algebra}
We assume that readers are familiar with basic (abstract) linear algebra (see Appendix for a quick review). We write $\bC$ for the set of complex numbers, $\imath$ for the imaginary of $\bC$ and $\overline{c}$ for the conjugate of $c\in\bC$, $U,V$ for linear spaces and $\bu,\bv,\bw$ for vectors in a linear space.

We start with the definition of tensor products.
\begin{defn}[Tensor product]
The tensor product space of $U$ and $V$ might be defined\footnote{There are alternative ways to define the tensor product up to an isomorphism.} as $U\otimes V\triangleq\spans\{(\bu_i,\bv_j)\}$ where $\{\bu_i\}$ and $\{\bv_j\}$ are bases of $U$ and $V$ respectively (see appendix for the definition of spans).
For any $\bu = \sum_ia_i\bu_i\in U$ and $\bv = \sum_jb_j\bv_j\in V$, their tensor product is defined as:
$$\bu\otimes\bv\triangleq\sum_i\sum_ja_ib_j(\bu_i,\bv_j)\in U\otimes V.$$
\end{defn}
Next, we consider linear maps. Let $\cL(U; V)$ denote the set of all linear maps from $U$ to $V$, and $\cL(U) \triangleq \cL(U; U)$ for all linear maps on $U$. $\cL(U;V)$ forms a linear space with addition $f+g : \bu \mapsto f(\bu)+g(\bu)$ and scalar multiplication $af : \bu \mapsto af(\bu)$. Note that identity map $I : \bu \mapsto \bu$ is a linear map and thus belongs to $\cL(U)$. Given two linear maps $f\in\cL(U;V)$ and $g\in\cL(W,U)$, let $f\circ g : \bw\mapsto f(g(\bw))$ be the composition of $f$ and $g$ which is again a linear map $f\circ g\in\cL(W,V)$; composition is associative, i.e., $f\circ (g\circ h) = (f\circ g)\circ h$. We can also define tensor products of sets of linear maps: since $\cL(U_1,V_1)$ and $\cL(U_2,V_2)$ are both linear spaces, we can define the tensor product space $\cL(U_1,V_1)\otimes\cL(U_2,V_2)\simeq\cL(U_1\otimes U_2; V_1\otimes V_2)$ as well as the tensor product of linear maps, i.e., $f\otimes g : \bu\otimes\bv \mapsto f(\bu)\otimes g(\bv) \in \cL(U_1\otimes U_2; V_1\otimes V_2)$.

We now turn to Hilbert spaces. They play an essential role in quantum mechanics.
\begin{defn}[Finite-dimensional Hilbert space over $\mathbb{C}$]
    A finite-dimensional Hilbert space $\cH$ over $\mathbb{C}$ (Hilbert space for short) is a finite-dimensional linear space over $\mathbb{C}$ equipped with an inner product $\<\cdot, \cdot\>$ mapping each pair of vectors to $\mathbb{C}$ such that\footnote{Finite-dimensional linear space is complete w.r.t. any linear norm and thus we omit the condition of completeness.}:
    \begin{enumerate}
        \item(conjugate symmetric) $\<\bu,\bv\> = \overline{\<\bv,\bu\>}$ for all $\bu,\bv\in\cH$;
        \item(linear on the second argument) $\<\bu,a\bv+\bw\> = a\<\bu,\bv\>+\<\bu,\bw\>$ for all $a\in\mathbb{C}$ and $\bu,\bv,\bw\in\cH$;
        \item(positive definite) $\<\bu,\bu\>\ge 0$ for all $\bu\in\cH$ and ``='' holds if and only if $\bu = 0$.
    \end{enumerate}
\end{defn}
Hilbert spaces are linear spaces and thus inherit all their definitions and results; e.g., $\cL(\cH)$ is the set of linear maps on $\cH$. We often call $f\in \cL(\cH_1;\cH_2)$ a (linear) operator. We can further define: 1). (induced) norm : $\|\bu\|\triangleq\sqrt{\<\bu,\bu\>}$. We call $\bu$ a unit vector if $\|\bu\|=1$; 2). orthonormal basis (ONB): a basis $\{\bu_i\}$ such that $\|\bu_i\| = 1$ and $\<\bu_i,\bu_j\>= 0$ if $i\neq j$; 3). outer product: $[\bu,\bv] : \bw\mapsto \<\bv,\bw\>\bu \in \cL(\cH_2;\cH_1)$ for $\bu\in\cH_1$ and $\bv\in\cH_2$; 4). adjoint of $A\in\cL(\cH_1,\cH_2)$ : the unique operator $A^\dag\in\cL(\cH_2,\cH_1)$ such that $\<\bu,A(\bv)\> = \<A^\dag(\bu),\bv\>$ $\forall\,\bu,\bv\in\cH$; 5). trace of $A\in\cL(\cH)$ : $\tr A\triangleq\sum_i\<\bu_i,A(\bu_i)\>$ where $\{\bu_i\}$ is an ONB of $\cH$. It is often convenient to view the partial trace of an operator in $\cL(\cH_1\otimes\cH_2)$ as a linear map, using tensor prodcuts; that is, we define the linear maps $\tr_1\in\cL(\cL(\cH_1\otimes\cH_2);\cL(\cH_2))$ as $\tr_1 : f \otimes g \mapsto \tr(g)f$ and $\tr_2\in\cL(\cL(\cH_1\otimes\cH_2);\cL(\cH_1))$ as $\tr_2 : f \otimes g \mapsto \tr(f)g$.

There are many important properties of operators $A\in\cL(\cH)$:
\begin{itemize}
    \item Hermitian: $A^\dag = A$;
    \item positive-semidefinite (or simply positive) (denoted by $\cL^+(\cH)$): $\forall\bu\in\cH, \<\bu,A(\bu)\>\ge 0$;
    \item density operator (denoted by $\cD^1(\cH)$): $A$ is positive and $\tr A = 1$;
    \item partial density operator (denoted by $\cD(\cH)$): $A$ is positive and $\tr A \le 1$;
    \item effect (or quantum predicate; 
    denoted by $\cO(\cH)$): $A$ and $I - A$ are both positive;
    \item unitary operator (denoted by $\cU(\cH)$: $A^\dag\circ A = A\circ A^\dag = I$;
    \item projection (denoted by $\cP(\cH)$ : $A$ is positive and $A\circ A = A$.
\end{itemize}
The positivity property can be used to define the  L\"owner order $A\sqsubseteq B$  order on linear maps: for any $A,B\in\cL(\cH)$, $A\sqsubseteq B$ iff $B - A$ is
positive. Note that $(\cO(\cH),\sqsubseteq)$ forms a complete partial order (CPO).

One important class of linear operators are the so-called super-operators. A super-operator from $\cH_1$ to $\cH_2$ is an element of $\cSO(\cH_1;\cH_2)\triangleq\cL(\cL(\cH_1);\cL(\cH_2))$. By convention, we write $\cI$ for the identity super-operator (which in fact is $I$ on $\cL(\cH)$). Similarly, we write $\cSO(\cH)\triangleq\cL(\cL(\cH))$ for the set of super-operators on $\cH$. We remark that $\cSO(\cdot;\cdot)$ is again an linear space; and thus we can define their tensor product. There are several important properties of super-operators $\cE$:
\begin{itemize}
\label{def-super-operator-property}
    \item positive: $\forall\,A\in\cL^+(\cH_1)$, $\cE(A)\in\cL^+(\cH_1)$;
    \item completely-positive (denoted by $\cCP(\cH_1;\cH_2)$): $\cE\otimes\cI$ is positive for identity super-operators $\cI\in\cSO(\cH^\prime)$ on any $\cH^\prime$;
    \item trace-preserving: $\forall\,A\in\cL^+(\cH_1)$, $\tr(\cE(A)) = \tr(A)$;
    \item trace-nonincreasing: $\forall\,A\in\cL^+(\cH_1)$, $\tr(\cE(A)) \le \tr(A)$;
    \item quantum operation (denoted by $\cQO(\cH_1;\cH_2)$) : completely-positive and trace-nonincreasing;
\end{itemize}

\noindent\emph{Linear algebraic structure.} Given a Hilbert space $\cH$, then all $\cH$, $\cL(\cH)$ and $\cSO(\cH)$ have linear algebraic structure, and thus we can freely use addition $+$ and scalar multiplication and apply the theory of linear space. However, the subsets such as effects (i.e. quantum predicates)  $\cO(\cH)$ and $\cQO(\cH)$ do not have such structure, i.e., they are not closed under addition or scalar multiplication; we need to be careful, for example, additional proofs are required to ensure $O_1+O_2\in\cO(\cH)$ even if $O_1,O_2\in\cO(\cH)$.

\subsection{Introduction to quantum mechanics}
\label{sec-quantum-mechanis}
Let us briefly introduce the fundamental concepts of quantum mechanics and their corresponding mathematical objects  in linear algebra. For more detail, please refer to the Appendix.

\emph{State space.} The state space of any isolated physical system is a Hilbert space $\cH$.
The state of a system is completely described by a unit vector $\bu\in\cH$.
To model the probabilistic ensembles of quantum states, \emph{(partial) density operators} are introduced; in detail, if the system is in one of the $\{\bu_i\}$ with probability $p_i$ ($\sum_i p_i=1$), then the system is fully characterized by $\rho\triangleq\sum_ip_i[\bu_i,\bu_i]\in\cD^1(\cH)$ the summation of outer product of $\bu_i$ with weight $p_i$.

\emph{Evolution.} The evolution of a closed quantum system is described by a unitary operator\footnote{Such transformation is fully determined by Schr\"odinger equation of the system.}. That is, if the initial state is $\bu$ and the evolution is described by $U$, then the final state is $U(\bu)$.
More generally, the evolution of a open quantum system which might interact with the environment, is modelled as a quantum operation $\cE\in\cQO(\cH)$, i.e., a complete-positive trace-nonincreasing superoperator on $\cH$; if the initial state is $\rho\in\cD(\cH)$, then the final state is $\cE(\rho)$.

\emph{Quantum measurement.} 
A quantum measurement on a system with state space $\cH$ is described by a collection of operators $\{M_m\}_{m\in J}$ ($M_m\in\cL(\cH)$ and $J$ is the index set)
such that $\sum_{m\in J}M_m^\dag\circ M_m = I$, where $J$ is the set of all possible measurement outcomes and $I$ the identity operator. It is physically interpreted as : if the state is $\bu$ immediately before the measurement then the probability that result $m$ occurs is $p(m) = \|M_m(\bu)\|$ and the post-measurement state is $\frac{M_m(\bu)}{\sqrt{p(m)}}$. If the state is described by density operator $\rho$, then $p(m) = \tr(M_m^\dag\circ M_m\circ \rho)$ and the post-measurement state is $\frac{M_m\circ \rho\circ M_m^\dag}{\tr(M_m^\dag\circ M_m\circ \rho)}$.
Performing quantum measurement is the only way to extract classical information (i.e., outcome) from a  quantum system.

\emph{Quantum predicates.}  Quantum predicates are defined as physical observables $O\in\cO(\cH)$.  The motivation comes from~\cite{DP06}. Informally, we can build for every observable $O$ a projective measurement $\cM = \{P_\lambda\}$ where $\lambda$ ranges over the eigenvalues of $O$ and $P_\lambda\in\cP(\cH)$ such that $O = \sum_\lambda P_\lambda$. Then suppose that the system is in state $\rho\in\cD^1(\cH)$; after performing the measurement $\cM$, we have probability $\tr(P_\lambda\circ\rho)$ to obtain as outcome $\lambda$. Moreover, note that \emph{expectation}, i.e., average value of outcome, is computed as $\sum_\lambda\tr(P_\lambda\circ\rho)\lambda = \tr(O\circ\rho)$. Thus, $\tr(O\circ\rho)$ is the expectation of $O$ in state $\rho$. This expectation might be interpreted as the degree that quantum state $\rho$ satisfies quantum predicate $O$.

\emph{Composite system.} 
The state space of a composite quantum system is the tensor product of the state spaces of the component physical systems. Specifically, if system $i\in\{1,2,\cdots,n\}$ is prepared in $\bu_i\in\cH_i$, then the joint state of the total system is $\bigotimes_{i=1}^n\bu_i = \bu_1\otimes\bu_2\otimes\cdots\otimes\bu_n\in\bigotimes_{i=1}^n\cH_i$.

Given a density operator $\rho\in\cD(\cH_A\otimes\cH_B)$ of a composite system $AB$, we define $\rho|_A\triangleq \tr_2(\rho)\in\cD(\cH_A)$ and $\rho|_B\triangleq\tr_1(\rho)\in\cD(\cH_A)$ 
as the reduced density operators of each subsystem $A$ and $B$. Physically, the reduced density operator fully describes the state of  a subsystem if all other subsystems are discarded or ignored.

\begin{wrapfigure}{r}{5.5cm}
    \includegraphics[width=0.4\textwidth]{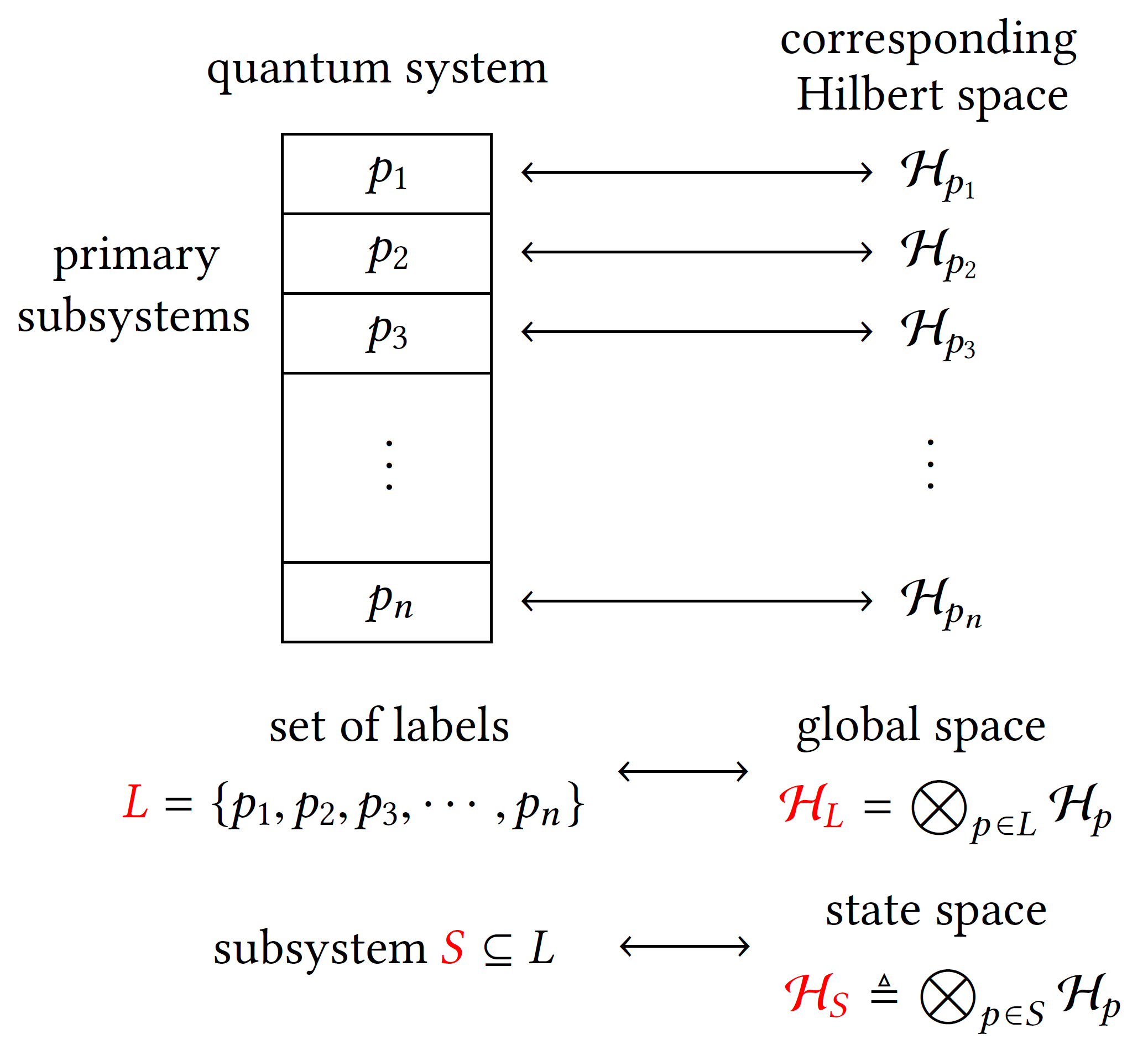}
    \caption{Global vs local state space}
    \label{fig-vars}
\end{wrapfigure}

\emph{Entanglement.} This is one of the most important features of the quantum world. Mathematically, it says that there exists state $\bw\in\cH_A\otimes\cH_B$ (a state of composite system $AB$) such that, there does not exist $\bu\in\cH_A, \bu\in\cH_B$ satisfies $\bw = \bu\otimes\bv$.
It suggests that it is impossible to preparing $\bw$ by preparing two independent states of $A$ and $B$ respectively; and what is more, the local operation in one of the system will influence the other, for example, the outcomes of performing measurement on $A$ and $B$ separately are correlated.

\section{Labelled Dirac Notation}
Dirac notation (a.k.a.\, bra-ket notation) is a widely used notation for representing quantum states. One of the main appeals of the notation is to simplify calculations
on quantum states. Informally, the crux of the Dirac notation is to provide a notation that can describe all mathematical entities and operations used in quantum mechanics and to interpret these ingredients uniformly as linear operators. By this means, the Dirac notation allows to calculate over descriptions of quantum systems using basic rewriting rules of linear algebra such as linearity, associativity and commutativity.

The standard Dirac notation combines the following ingredients:
\begin{enumerate}
\item ket $|\cdot\>$ represents a state, i.e., $|u\>\triangleq\bu\in\cH$.
  Since $\bC$ itself is a Hilbert space with $\<a,b\> \triangleq \overline{a}b$,
  states can also be viewed as elements of $\cL(\bC;\cH)$;
\item bra $\<\cdot|$ represents a co-state, i.e, a linear map defined by $\<u|\triangleq \bv \mapsto \<\bu,\bv\>\in\cL(\cH;\bC)$;
  
\item inner product $\<\cdot|\cdot\>$, i.e., $\<u|v\>\triangleq\<\bu,\bv\>\in\bC$; inner products can also be viewed as linear maps over $\bC$, and more specifically as the composition of the bra and the kets viewed as linear maps, i.e.\, $\<u|v\>\in\cL(\bC)$ with $\<u|v\> = \<u|\circ|v\>$;
  
\item outer product $|\cdot\>\<\cdot|$, i.e., $|u\>\<v|\triangleq[\bu,\bv]\in\cL(\cH_1;\cH_2)$ if $\bu\in\cH_2$ and $\bv\in\cH_1$; this definition of outer product is also consistent with composition $|u\>\<v| = |u\>\circ\<v|$;
\item tensor product $|\cdot\>|\cdot\>$ and $\<\cdot|\<\cdot|$, i.e., $|u\>|v\>\triangleq\bu\otimes\bv\in\cH_1\otimes\cH_2$ and $\<u|\<v|\triangleq\<u|\otimes\<v|\in\cL(\cH_1\otimes\cH_2;\bC)$ if $\bu\in\cH_1$ and $\bv\in\cH_2$.
\end{enumerate}

\begin{example}
The Dirac notation eases equational reasoning, as shown by the following example (the linear map view of expressions is shown in {\color{blue}{blue}}).
\begin{align*}
&\quad\ (|\phi\>\<\psi|)(|\alpha\>\<\beta|) \xlongequal{\ \ \ \, assoc.\ \ \ \, } |\phi\>(\<\psi|\alpha\>)\<\beta|
\xlongequal{\ \ scalar\ \ }(\<\psi|\alpha\>)(|\phi\>\<\beta|)\\
&{\color{blue}(|\phi\>\circ\<\psi|)\circ(|\alpha\>\circ\<\beta|)\qquad
|\phi\>\circ(\<\psi|\circ|\alpha\>)\circ\<\beta|\qquad\ 
(\<\psi|\alpha\>)(|\phi\>\circ\<\beta|)}
\end{align*}
\end{example}

Labelled Dirac notation is an enhancement of Dirac notation that uses subscripts to identify the subsystems where a ket/bra/etc lies or operates; for example, $|\phi\>_S, {}_S\<\psi|$, ${}_S \< \phi | \psi \>_S$ and $|\phi\>_S\<\psi|$ range over states, co-states, inner products and outer products on subsystem $S$, where $S$ is a set of labels drawn from some fixed global set $L$. This notation has two benefits: first, it makes it possible to describe subsystems of a quantum system, without the need of lifting the subsystem to the global state by means of tensor products. This makes the description of quantum subsystems much more concise and easier to manipulate in calculations. Second, it exposes identities that are commonly used in calculation. One such identity is commutativity of tensor product, i.e.\, $|\phi\>_S|\psi\>_{S^\prime} = |\psi\>_{S^\prime}|\phi\>_S$ provided $S$ and $S^\prime$ are disjoint.

Expressions based on labelled Dirac notation admit a local interpretation. Specifically,  we can define in addition to the global state space $\cH_S = \bigotimes_{p\in S}\cH_p$ for every $S\subseteq L$; see Figure~\ref{fig-vars}.  Then one can interpret expressions based on labelled Dirac notations as linear maps between two local state spaces. This interpretation uses lifting to extend operations from some subspace to a cylindric extension of that space.
\begin{defn}[Cylindrical Extension]
\label{def-cylindric}
Given two subsets $S\subseteq T\subseteq L$ and a linear operator $A\in\cL(\cH_S)$, we define the cylindrical extension of $A_S$ in $T$ as $\cl_{T}(A) \triangleq A_S \otimes I_{T\backslash S}.$ If $T = L$, we simply write $\cl(A)$ for $\cl_L(A)$.
\end{defn}
To illustrate the use of cylyndrical extensions, suppose that we have two disjoint subsystems $S_1$ and $S_2$ with initial state $|\phi\>_{S_1}|\psi\>_{S_2}$, and we apply the unitary transformation $U_{S_1}$ to $S_1$. To model the result of this computation, one can first lift $U_{S_1}$ to $\cl_{S_1\cup S_2}(U_{S_1}) = U_{S_1}\otimes I_{S_2}$ (operator on $\cH_{S_1\cup S_2}$) and then apply the resulting map to the initial state:
$$(U_{S_1}\otimes I_{S_2})(|\phi\>_{S_1}|\psi\>_{S_2}) = (U_{S_1}|\phi\>_{S_1})(I_{S_2}|\psi\>_{S_2}) = (U_{S_1}|\phi\>_{S_1})|\psi\>_{S_2}.$$
However, since we have already labelled all subsystems in the formula, such lifting step can be automatically identified via the context, and thus, we might simply write:
\begin{equation}
\label{eqn-dirac-local-example}
    U_{S_1}(|\phi\>_{S_1}|\psi\>_{S_2}) = (U_{S_1}|\phi\>_{S_1})|\psi\>_{S_2}
  \end{equation}
  This notation captures the intuition that we apply $U_{S_1}$ locally to $|\phi\>_{S_1}$ while keeping $|\psi\>_{S_2}$ unchanged. Our interpretation of labelled Dirac notation
obeys the principle of \lq\lq\emph{localizing objects as much as possible}\rq\rq.

\begin{example}
Suppose $S$ and $T$ are two disjoint subsystems with orthonormal basis $\{|v_i\>_S\}_{i\in J}$ and $\{|u_i\>_T\}_{i\in J}$ respectively. We define the injection of a function $A : J\times J\ra \bC$ to a linear operator on $S$ (or, $T$) by $A[S] \triangleq A(i,j)|v_i\>_S\<v_i|$ (or, $A[T]\triangleq A(i,j)|u_i\>_T\<u_i|$). Let 
$|\Phi\>$ be the maximally entangled state on $S$ and $T$, i.e., $|\Phi\> = \sum_{i\in J}|v_i\>_S|u_i\>_T$. Then, for any $A$, applying $A[S]$ on $|\Phi\>$ yields the same states if we perform $A^T[T]$ on $|\Phi\>$ where $A^T(i,j) = A(j,i)$. This can be proved using labelled Dirac notation:
\begin{align*}
	A[S]|\Phi\> &= \sum_{m n}A(m,n)|v_m\>_{\color{blue}S}{\color{red}\<v_n|}\sum_{i}{\color{red}|v_i\>_{\color{blue}S}}|u_i\>_{T} 
	= \sum_{m n i}A(m,n){\color{red}{}_{\color{blue}S}\<v_n|v_i\>_{\color{blue}S}}|v_m\>_{S}|u_i\>_{T} \\
	&= \sum_{m i}A(m,i)|v_m\>_{S}|u_i\>_{T}
	= \sum_{m i j}A(j,i){\color{red}{}_{\color{blue}T}\<u_j|u_m\>_{T}}|v_m\>_{S}{\color{green}|u_i\>_{\color{blue}T}} \\
	&= \sum_{m i j}A(j,i){\color{green}|u_i\>_{\color{blue}T}}{\color{red}\<u_j|}\big({\color{red}|u_m\>_{T}}|m\>_{S}\big)
	= \sum_{i j}A(j,i)|u_i\>_{T}\<u_j|\sum_{m}|v_m\>_{S}|u_m\>_{T} \\
	&= A^T[T]|\Phi\>
\end{align*}
This derivation showcases the benefits of labelled Dirac notation : with the help of blue colored labels, we can quickly identify which (co-)states can be composition and rearranged (using associativity or scalar property) without worrying about the order of tensor product (such as in the third line). Furthermore, note that in the first line, we do not need to write out the lifting details ($A[S]$ should be lifted to $A[S]\otimes I_T$).
Instead we directly identify that $|v_m\>_{\color{blue}S}{\color{red}\<v_n|}$ will act on ${\color{red}|v_i\>_{\color{blue}S}}$ while leaving $|u_i\>_{T}$ unchanged.
\end{example}


\section{Syntax and semantics of \textbf{qwhile}}\label{sec:qwhile}

In this section, we present the syntax and denotational semantics of \textbf{qwhile},
a core language that follows the classic control and quantum data paradigm. 

\subsection{Syntax}\label{sec-abstract-syntax}
We first define an abstract syntax of \textbf{qwhile} programs. The specificity of this syntax is that it does not use variables. Later, we show how to instantiate the language and its semantics to programs with named variables.
\begin{defn}[Abstract syntax]
Quantum programs are generated by the following syntax:
\begin{align}
    C ::= &\aabort\ |\ \askip\ |\ C_1;C_2\ |\ \ainit{\rho_S} \ |\ 
    \aapply{U_S} \ | \nonumber\\
    &\acond{m}{\cM_S}{C}\ |\ \awhile{\cM'_S}{b}{C}
\end{align}
where $b$ is a boolean value, $S$ ranges over subsets of a set $L$ of labels, $\rho_S$, $U_S$, $\cM_S$ and $\cM'_S$  range respectively over density operators, unitary operators, quantum measurements on $\cH_S$, and two-value measurements (i.e., $\cM'_S = \{M_{0}, M_{1}\}$).
\end{defn}
The initialization $\ainit{\rho_S}$ sets the subsystem $S$ to quantum state $\rho_S$. The unitary transformation $\aapply{U_S}$ apply $U_S$ on subsystem $S$. The $\kif$-statement $\acond{m}{\cM_S}{C}$ first performs quantum measurement $\cM_S = \{M_{m}\}$ on subsystem $S$ and then executes the subprogram $C_m$ according to the outcome $m$. For the $\kwhile$-statement $\awhile{\cM_S'}{b}{C}$, if the outcome of two-value measurement $\cM_S'$ on subsystem $S$ is $\neg b$ then the program terminates; otherwise, the program executes subprogram $C$ and then repeats the loop again.

For convenience, we further define the syntactic sugar for sequential programs:
\begin{equation}
    \afor{i}{\ \leftarrow J}{C} \equiv C_{i_0};C_{i_1};\cdots;C_{i_n}
\end{equation}
where $J = \{i_0,i_1,\cdots,i_n\}$ is a sequence of indexes. We simply write $\afor{i}{\ <n}{C}$ if $i$ ranges over $0,1\cdots,n-1$.

\subsection{Semantics}
\label{sec-semantics}

The denotational semantics of programs is defined inductively on their structure.  The semantics of loops is based on the (syntactic) notion of approximation.  \begin{defn}[Syntactic Approximation of While; c.f. \cite{Ying11}] For a given quantum loop $\awhile{\cM_S}{b}{C}$, we inductively define its $k$-th syntactic approximation $$\akwhile{\cM_S}{b}{C}{k}$$ for any integer $k\ge 0$ as follows:
  \begin{equation*}
    \left\{
    \begin{array}{lll}
        \akwhile{\cM_S}{b}{C}{0} & \equiv&\aabort \\
        \akwhile{\cM_S}{b}{C}{k+1} & \equiv&\kif\ \cM_S = \neg b\ \rightarrow\ \askip \\
        &&\square\ \qquad\quad\! b\ \ \ \rightarrow\ C;\akwhile{\cM_S}{b}{C}{k}\ \kfi
    \end{array}
    \right.
\end{equation*}
\end{defn}
Informally, the semantics of a loop is the limit of the semantics of its syntactic approximations. This is captured by the following definition.
\begin{defn}[Semantics]
The semantics $\sem{C}$ of a quantum program $C$ is a super-operator on $\cH_L$ (that is, linear map from $\cL(\cH_L)$ to $\cL(\cH_L)$) inductively defined as follows:
\begin{enumerate}
    \item $\sem{\aabort} \triangleq 0$; i.e., $\sem{\aabort} : \rho\mapsto 0$;
    \item $\sem{\askip} \triangleq \cI_L$; i.e., $\sem{\askip} : \rho\mapsto \rho$;
    \item $\sem{C_1;C_2} \triangleq \sem{C_1}\circ\sem{C_2}$; i.e., $\sem{C_1;C_2} : \rho_L\mapsto \sem{C_1}(\sem{C_2}(\rho))$;
    \item $\sem{\ainit{\rho_S}} : \rho_L \mapsto \tr_S(\rho)\otimes\rho_S$;
    \item $\sem{\aapply{U_S}} : \rho \mapsto U_S\rho U_S^\dag$;
    \item $\sem{\acond{m}{\cM_S}{C}} : \rho_L \mapsto \sum_m\sem{C_m}(M_{m}\rho M_{m}^\dag)$;
    \item $\sem{\awhile{\cM_S}{b}{C}} \triangleq \lim_{k\rightarrow\infty}\sem{\akwhile{\cM_S}{b}{C}{k}}$.
    \end{enumerate}
 Note that we use  labelled Dirac notation for defining the semantics of unitary transformations; in detail, $U_S\rho U_S^\dag \triangleq \cl(U_S)\circ\rho\circ\cl(U_S^\dag)$. We also use labelled Dirac notation $M_m\rho M_m^\dag$ for  defining the semantics ofconditionals and loops.
\end{defn}
Note that the existence of limit in (7) always exists according to monotone convergence theorems on ordered Hilbert spaces---where the order is L\"owner order. Note that our semantics contrasts with classic domain-theoretical semantics which interpret loops as a least fixed point over the complete partial order (CPO) of quantum operations (recall that a quantum operation is completely-positive and trace-nonincreasing super-operator). However, one can show that $\sem{C}$ is a quantum operation, and that our semantics coincides with the domain-theoretic semantics. The proof of equivalence is based on the following lemma, which states that supremum in the CPO of quantum operations is consistent with topological limits of super-operators:
$$\forall\,\text{non-decreasing sequence}\ \cE_i\in\cQO(\cH),\ \bigsqcup_i\cE_i = \lim_{i\rightarrow\infty}\cE_i.$$
Note that the choice to define the semantics of programs as super-operators rather than quantum operation is motivated by the fact that super-operators enjoy linear algebraic properties that are not verified by quantum operations---see Section~\ref{sec:prelim}.

A key aspect of our semantics is that it naturally induces a local semantics over the subsystem defined by its variables. This is captured by the following lemma, which
is an adaptation of \cite[Proposition 3.3.5]{Ying16} to our setting.
\begin{lemma}[Localization of semantics]\label{lem:loc:sem}
Suppose $C$ is a quantum program, and let $\pset{C}$ be the union of all subsystems mentioned in $C$. There uniquely exists a local super-operator $\lsem{C} : \cL(\cH_{\pset{C}}) \rightarrow \cL(\cH_{\pset{C}})$ such that $\sem{C} = \lsem{C}\otimes\cI_{\overline{\pset{C}}}$ where $\cI_{\overline{\pset{C}}}$ is the identity super-operator on subsystem $\overline{\pset{C}}\triangleq L\backslash \pset{C}$. We call $\lsem{C}$ the local semantics of $C$.
\end{lemma}
This localization property plays an essential role in the proof of soundness of Hoare logic.
It states that the program $C$ will influence at most those subsystems entangled with $\pset{C}$, while leave those uncorrelated subsystems unchanged.

\subsection{Concrete syntax}
\label{sec-variable-type}
Although the abstract syntax suffices to develop the meta-theory of \textbf{qwhile}, the formalization of textbook algorithms and their correctness proofs is greatly simplified by switching to a concrete syntax with support for typed variables. In order to maximize convenience, the introduction of variables must satisfy two desiderata: first, variables should be given arbitrary types, rather than being limited to represent qubits. Second, variables should be closed under cartesian products and projections,
so that it is sufficient to use the following (selected) syntax:
$$\mx := |\mt\>\ \big|\ \mx := U[\mx]\ \big|\ \icond{\mt}{\rm meas}{\mx}{C}\ \big|\ \iwhile{\rm meas}{\mx}{b}{C}$$
for initialization, unitary transformation, if and while statements using computational basis measurement as guards, respectively. 

Instantiating the semantics of abstract programs to concrete programs requires several steps:
\begin{itemize}
\item define the Hilbert space associated to a (finite) set with its elements as computational basis;

\item model quantum variables: a quantum variable $\mx$ of type $\mT$ associates its symbolic name and a storage location (quantum subsystem $\pset{\mx}$), together with the mapping between associated Hilbert space of $\mT$ and state space of its quantum subsystem.
We write $|u\>_\mx\in\cH_{\pset{\mx}}$ and $A[\mx]\in\cL(\cH_{\pset{\mx}})$ for the injection of 
$\bu\in\cH_\mT$ and $A\in\cL(\cH_\mT)$ respectively.

\item model composition of quantum variables. Suppose two quantum variables $\mx_1$ of type $\mT_1$ 
and $\mx_2$ of type $\mT_2$ with disjoint domains (i.e., $\pset{\mx_1}\cap\pset{\mx_2} = \emptyset$), we manipulate the pair \coqe{$\overline{\mx} \triangleq $[x$_1$,x$_2$]} as a quantum variable of type \coqe{T$_1\ast$T$_2$} which consistent with actions on its components.

\end{itemize}

\section{Program logic}
CoqQ implements a program logic that is proved sound with respect to the denotational semantics of \textbf{qwhile} programs.

\subsection{Judgments}
\label{sec-correctness}

Correctness of \textbf{qwhile} programs is expressed using Hoare triples. A (mild) novelty of our approach is that we allow assertions to be arbitrary linear operators rather than observables. This simplifies the proof rules, since linear operators have better algebraic properties than operators. Of course, our notion of validity coincides with the usual notion of validity when the pre- and post-conditions are observables.
\begin{defn}[Valid judgment]
\label{def-local-correct}
A Hoare triple is a judgment of the form $\{A_{S_1}\}C\{B_{S_2}\}$ with
$S_1,S_2\subseteq L$ and $A\in\cL(\cH_{S_1}),\ B\in\cL(\cH_{S_2})$. The validity of a Hoare triple is defined as follows (recall Definition \ref{def-cylindric} for cylindrical extension): 
\begin{itemize}
    \item total correctness: $\models_{\rm t} \{A_{S_1}\}C\{B_{S_2}\}$ if for all $\rho\in\cD(\cH_L)$, 
    \begin{equation}
    \label{eqn-total-local}
        \tr[\cl(A_{S_1})\circ\rho]\leq \tr[\cl(B_{S_2})\circ\sem{C}(\rho)];
    \end{equation}
    \item partial correctness: $\models_{\rm p} \{A_{S_1}\}C\{B_{S_2}\}$ if for all $\rho\in\cD(\cH_L)$, 
    \begin{equation}
    \label{eqn-partial-local}
        \tr[\cl(A_{S_1})\circ\rho]\leq \tr[\cl(B_{S_2})\circ\sem{C}(\rho)]+[\tr(\rho)-\tr(\sem{C}(\rho)].
    \end{equation}
\end{itemize}
\end{defn}
Recall that $\tr[\cl(A_S)\circ\rho] = \tr(A_S\circ\rho|_S)$ is the expectation of $A_S$ in local state of $\rho$ in subsystem $S$. Therefore, $\models_{\rm t} \{A_{S_1}\}C\{B_{S_2}\}$ says that the expectation of $A_{S_1}$ in local input of subsystem $S_1$ is smaller than (equals to for $\models_{\rm t}^{\rm s} \{A_{S_1}\}C\{B_{S_2}\}$) the expectation of $B_{S_2}$ in local output of subsystem $S_2$. Validity of partial correctness statements is stated similarly, but the probability of non-termination $\tr(\rho)-\tr(\sem{C}(\rho)$ is added to the rhs of the inequality.

In some circumstances, it is desirable to establish a stronger notion of correctness, where the inequalities in Eqn. (\ref{eqn-total-local}) or (\ref{eqn-partial-local}) are replaced by =. We use the superscript ``${\rm s}$'' (for saturated) to indicate that the equality holds, e.g., $\models_{\rm p}^{\rm s}$ or $\models_{\rm t}^{\rm s}$. We remark that deriving saturation brings stronger results, i.e., the equivalence (rather than inequlity) of pre and post-expectation, as used in the HSP algorithm, while require no extra effort since most rules apply to both saturated and non-saturated cases.

Finally, it is often convenient to use a special form of judgments $\{AA^\dag\}C\{BB^\dag\}$, for which we introduce the following syntactic sugar (we use the color \textcolor{purple}{purple} rather than introducing new notation):
$$\cgrule{}{}{A}{C}{B}$$
This notation is particularly useful when both pre- and post-conditions are expressed as states in labelled Dirac notation. In this case, $\cgrule{}{}{|u\>_{S_u}}{C}{|v\>_{S_v}}$ holds whenever the program $C$ transforms the input state $|u\>_{S_u}$ into the output state $|v\>_{S_v}$---under the proviso that $\||u\>_{S_u}\| = \||v\>_{S_v}\| = 1$.
This provides a \emph{human-readable} statement that looks like textbook statements of program correctness. This readability extends not only to statements but also to proofs (see Fig. \ref{fig-hsp-proof} for an example). In particular, using labelled Dirac notation as pre- and post-conditions allows us to write, interpret, derive and calculate assertions locally, using the equational theory attached to the notation.

\subsection{Inference rules}
\label{sec-inference-rule}
Our proof system provides a rich set of inference rules for reasoning about valid Hoare triples. Proof rules are divided into several categories: syntax-directed rules, which are specific to one program construct, including custom rules for \textbf{for} loops;  structural rules, which can be interspersed with syntax-directed rules to ease/enable reasoning;
state-based rules, which are based on the representation of assertions as states in labelled Dirac notation. Rules from the first and second categories are inspired from existing literature. However, rules from the third category, notably the rule (R.Inner), seem new.  Fig. \ref{fig-rule} presents a selection of forward rules for partial correctness based on the concrete syntax. Additional rules for total correctness, abstract syntax and backward reasoning are given in the appendix.

\begin{figure}[t]\centering
\begin{equation*}\begin{split}
&({\rm Ax.Sk})\ \models_{\rm p}\{A\}\mathbf{Skip}\{A\} \qquad 
 ({\rm Ax.UTF})\ \models_{\rm p}\{A\}\mx := U[\mx]\{U[\mx]AU[\mx]^{\dag}\}\\
&({\rm Ax.InF}) \ \frac{S\cap\pset{x} = \emptyset}{\models_{\rm p}\{A_S\}\mx := |\mt\>\{A_S\otimes|\mt\>_\mx\<\mt|\}} \qquad 
({\rm R.SC})\ \frac{\models_{\rm p}\{A\}S_1\{B\}\quad\ \ \models_{\rm p}\{B\}S_2\{C\}}{\models_{\rm p}\{A\}S_1;S_2\{C\}}\\
&({\rm R.IF})\ 
\frac{\models_{\rm p}\{A_\mt\}C_\mt\{B\}\ {\rm for\ all}\ \mt}{\models_{\rm p}\big\{\sum_{\mt : \mT}
|\mt\>_\mx\<\mt|A_m|\mt\>_\mx\<\mt|\big\}\icond{\mt}{\rm meas}{\mx}{C}\{B\}}\\
&({\rm R.LP.P})\ 
\frac{R := |b\>_\mx\<b|A|b\>_\mx\<b|+|\neg b\>_\mx\<\neg b|B|\neg b\>_\mx\<\neg b|\quad \models_{\rm p}\{A\}C\{R\}\quad A\sqsubseteq I\quad B\sqsubseteq I}{\models_{\rm p}\{R\}
  \iwhile{\rm meas}{\mx}{b}{C}\{B\}}\\
&({\rm R.PC.P})\ \frac{\begin{split}
		\forall i, \models_{\rm p}\{A_{i,S_{A_i}}\}P_i\{B_{i,S_{B_i}}\}\quad \forall i, 0\sqsubseteq A_{i,S_{A_i}}\sqsubseteq I_{S_{A_i}} \quad \forall i, 0\sqsubseteq B_{i,S_{B_i}}\sqsubseteq I_{S_{B_i}}\\
		\forall i\neq j, (\pset{C_i}\cup S_{A_i}\cup S_{B_i}) \cap (\pset{C_j}\cup S_{A_j}\cup S_{B_j}) = \emptyset\qquad\ \ 
\end{split}}{\models_{\rm p}\{\bigotimes_iA_{i,S_{A_i}}\}\mathbf{for}\ i\ \mathbf{do}\ C_i\{\bigotimes_iB_{i,S_{B_i}}\}}\\
&({\rm Ax.UTFP})\ \ \ \frac{\forall i\neq j, \pset{\mx_i} \cap \pset{\mx_j} = \emptyset}{\models_{\rm p}\{A\}\mathbf{for}\ i\ \mathbf{do}\ \mx_i := U_i[\mx_i]\{(\bigotimes_iU_i[\mx_i])A(\bigotimes_iU_i[\mx_i])^\dag\}}\\
&({\rm Ax.InFP})\ \ \ \frac{\forall i\neq j, \pset{\mx_i} \cap \pset{\mx_j} = \emptyset}{\models_{\rm p}\{1\}\mathbf{for}\ i\ \mathbf{do}\ \mx_i := |\mt_i\>\{\bigotimes_i|\mt_i\>_{\mx_i}\<\mt_i|\}}\\
&({\rm R.Or})\ \frac{A\sqsubseteq
A^{\prime}\ \ \models_{\rm p}\{A^{\prime}\}C\{B^{\prime}\}\ \ 
B^{\prime}\sqsubseteq B}{\models_{\rm p}\{A\}C\{B\}}\quad
({\rm R.CC.P})\ \frac{\forall i, \models_{\rm p}\{A_i\}C\{B_i\}\ \ \forall i, 0\le\lambda_i\ \ \sum_i\lambda_i\le 1}{\models_{\rm p}\{\sum_i\lambda_i A_i\}C\{\sum_i\lambda_i B_i\}}\\
&({\rm Ax.Inv})\ \frac{A_S\sqsubseteq I_S\quad S\cap \pset{C} = \emptyset}{\models_{\rm p}\{A_S\}C\{A_S\}}\qquad ({\rm R.El})\ \frac{\models_{\rm p}\{A_{S_A}\}C\{B\}\quad S_A\cap S = \emptyset}{\models_{\rm p}\{A_{S_A}\otimes I_{S}\}C\{B\}}\\
&({\rm Frame.P})\ \frac{\models_{\rm p}\{A_{S_A}\}C\{B_{S_B}\}\quad 0\sqsubseteq R_S\sqsubseteq I_S\quad (\pset{C}\cup S_A\cup S_B)\cap S = \emptyset\quad}{\models_{\rm p}\{A_{S_A}\otimes R_{S}\}C\{B_{S_B}\otimes R_{S}\}}\\
&({\rm R.Inner})\ \frac{\models_{\rm t}^{\rm s}\{1\}C\{|v\>_{S_v}\<v|\}\quad \||v\>_{S_v}\|\le 1\quad S_u\subseteq S_v }{\models_{\rm t}^{\rm s}\{\|{}_{S_u}\<u|v\>_{S_v}\|^2\}C\{|u\>_{S_u}\<u|\}} \\
 & ({\rm Ax.UTF'})\ \ \ \cgrule{\rm p}{}{|v\>_S}{\mx := U[\mx]}{U[\mx]|v\>_S} \qquad ({\rm Ax.InF'}) \ \ \ \frac{S\cap\pset{x} = \emptyset}{\cgrule{\rm p}{}{|v\>_S}{\mx := |\mt\>}{|v\>_S\otimes|\mt\>_\mx}}
\end{split}
\end{equation*}
\caption{Selected inference rules provided in CoqQ.
In (R.Inner), $\models_{\rm t}^{\rm s}$ stands for the saturated total correctness, i.e., the inequality in Eqn. (\ref{eqn-total-local}) is replaced by ``=''.
}\label{fig-rule}
\end{figure}

\emph{Syntax-directed rules}.
The rules (Ax.Sk), (Ax.InF), (Ax.UTF), (R.SC), (R.IF) and (R.LP.P) are used to reason about the core constructs of the \textbf{qwhile} language. In addition, the rules (R.PC.T), (Ax.UTFP) and (Ax.InFP) are used to reason about \textbf{for} loops. programs.  These rules have the side-condition of ``disjointness'' -- assume that the programs (predicates) are about different quantum subsystems.

\emph{Structural rules}. Structural rules are not tied to a specific language construct and can be interspersed with other rules to simplfy correctness proofs. Rule (R.CC.P) simplifies correctness proofs by enabling linear combinations of pre- and post-conditions. Rule (Ax.Inv) says that is any predicate disjoint from the footprint
of the program is preserved. (R.EI) allows us to lift predicates to a larger space and can be regarded as using ``auxiliary variables''. Finally, (Frame.P) allows to focus on the part of the predicates that are related to the program while keeping the rest part unchanged.

\emph{State-based rules}. The rules ({\rm Ax.UTF'}) and ({\rm Ax.InF'}) are state-based variants of the rules (Ax.UTF) and (Ax.InF).  Besides the human-readability, these rules reduce the calculations on labelled Dirac notation by half; for example, compare to the rule (Ax.UTF); in this case, the post-condition is ${\color{blue}(U[\mx]|v\>_S)}{\color{red}({}_S\<v|U[\mx]^\dag)}$; note that we do not need to calculate the red part since it is just the adjoint of the blue part.  Most of the rules have such variants (which we label their name by a single quote); we do not display them here due to space limitations.

The new rule (R.Inner) provides a way to link forward and backward reasoning, i.e., it derives the precondition for a given post-condition from a judgment derived by forward reasoning. Forward reasoning is more suitable for programs with fixed inputs (or starting with initialization), and is usually intuitive and relatively simple; backward reasoning is suitable when postcondition is given, but with a possibly much more involved computation. HSP algorithm is such an example; see Section~\ref{sec:hsp}.

\subsection{Soundness and weakest precondition}

All inference rules are formalized as lemmas, and proved from first principles. Formally, we claim:
\begin{theorem}[Soundness]
All the inference rules displayed in Section \ref{sec-inference-rule} are sound with respect to the judgment defined in Definition \ref{def-local-correct}.
\end{theorem}
In addition to the soundness proof, we show that for all observables in $\cO(\cH_L)$, the weakest (liberal) precondition is well-defined, i.e., and satisfies $\models\{A\}C\{B\}$ if and only if $A\sqsubseteq wp.C.B$ for total correctness (or $wlp.C.B$ for partial correctness).

\section{Implementation}

All the aforementioned concepts have been formalized in the
Coq~\cite{coq} proof assistant. Our formalization relies on the
MathComp~\cite{mathcomp} library for basic data structures such as
finite sets, ordered fields, linear algebra, and vector spaces, and on
the MathComp Analysis~\cite{mathcomp-analysis} library, an ongoing
attempt at providing a library for classical analysis in Coq that is
compatible with the mathematical objects introduced in the MathComp
library.

\subsection{Mathematical libraries}

\noindent \textbf{Finite Dimensional Hilbert Spaces.}
Although the MathComp library has an extensive theory about finite
dimensional linear spaces and linear algebra, nothing is said when it
comes to Hilbert spaces. Hence, we extended the MathComp hierarchy with
a new type \coqe{hermitianType R}, for \coqe{R} a real domain, that
extends the MathComp defined type \coqe{vectType R[i]} of finite
dimensional linear spaces over the complex closure of \coqe{R}.
This new structure comes with an hermitian inner product
\coqe{[< . , . >]}, i.e. a sesquilinear form over \coqe{R[i]}.
We also formalized the core theory of Hermitian spaces. Notably,
we defined and proved correct the Gram–Schmidt process that allows
the orthonormalization a set of vectors w.r.t. an inner product.
Last, our library comes with a sub-type \coqe{chsType R}, of
\coqe{hermitianType R}, that denotes hermitian spaces that comes with
an orthonormal canonical basis.

\smallskip

\noindent \textbf{Tensor Product \& Hilbert Spaces.}
Our library comes with a definition of the tensor product of linear
spaces. However, instead of defining the tensor product of \emph{two}
linear spaces only, we give a direct definition for the tensor product of
a finite family of linear spaces. (The definition is ubiquitous as,
for a  fixed field $k$ and up to linear spaces isomorphism, the tensor
product operator forms a commutative monoid over $k$-linear spaces)
Doing so allows us to have a unique type for the isomorphic spaces
$(E \otimes F) \otimes G$ and $E \otimes (F \otimes G)$, lowering
the use and tedious handling of explicit linear spaces isomorphisms.
Our formal definition of tensor products relies on the basis-based
definition of tensor products: if $\{E_i\}_i$ is a finite family 
of linear spaces over a field $k$ with respective basis
$\{ e_{i,j} \}_j$, then $\bigotimes_i E_i$ is any vector space
with the formal basis
$\{ e_{1, j_1}   \otimes
      e_{2, j_2} \otimes
      \cdots     \otimes
      e_{n, j_n} \}_{j_1, j_2, \ldots, j_n} .$

Related to tensor products, our library comes with a notion of
multi-linear maps over a finite family of linear spaces $\{E_i\}_i$
and a formal proof of the universal property of tensor products:
any multi-linear map over $\bigtimes_i E_i$ can be canonically lifted
to a linear map over $\bigotimes_i E_i$.

Last, we demonstrate that the tensor product of Hilbert spaces
is itself an Hilbert space.

\smallskip

\noindent \textbf{Labelled Dirac Notation.}  In principle, a labelled
Dirac notation is interpreted as a linear map from its domain to
codomain which is suitable to be implemented as a dependent
type. 
However, the tensor product and automatic lifting make its (co)domain
changes frequently during calculation and thus face the usability
problem caused by type cast -- a technical obstacle for dependent
type.  Our library come up with a definition of labelled Dirac
notation as a non-dependent type which is carefully designed to have
linear algebraic structure.  On the other hand, we use the
\emph{Canonical Mechanism} to trace the (co)domain of a labelled Dirac
notation.  We provide the canonical instance for each binary
operation/big operator of labelled Dirac notation, and thus Coq will
automatically infer its (co)domain from its structure.  A important
feature of our formalization is that it allows to use big operators
for tensor products, i.e.\, $\bigotimes_{i}$, since tensor products
have a commutative monoid structucture with $1$ as the identity 
of tensor product.

\smallskip
\noindent \textbf{Concepts for Quantum Framework}.
As instances of \coqm{vectType}, Hilbert space $\cH$, linear maps $\cL(\cH)$ and super-operator $\cSO(\cH)$ inherited the linear algebraic structure and form the basis of our quantum framework.  
Besides, we formalize most of the fundamental concepts in quantum mechanics using \coqe{Structure}, including normalized state (\coqm{NSType}), completely-positive maps (\coqm{CPType}), quantum operation (\coqm{QOType}), quantum channel (\coqm{QCType}), (partial) orthonormal basis (\coqm{PONBasis} and \coqm{ONBasis}), trace-nonincreasing maps (\coqm{TNType}), quantum measurement (\coqm{QMType/TPType}), and a series of subsets of linear operators -- hermitian, positive-semidefinite, observable, (partial) density operator, unitary, (rank-1) projection; see Fig. \ref{fig:hierachy_graph}.
A state/operator/super-operator/function with a proof of some property can be declared as canonical to the corresponding definition and then inherits the properties of that definition.

\smallskip 
\noindent \textbf{Other Results.}  We also prove several results that
are not covered by MathComp Analysis. These results include the
Bolzano-Weierstrass theorem on Euclidian spaces, the monotone
convergence theorem for finite-dimensional linear spaces. The latter
is directly applied to define the semantics of \textbf{qwhile}
programs. Finally, we develop a library for matrix norm and subspace
theory (represented by projection and canonical to ortho-modular
lattice).

\newcommand{\vsp}{.8}
\newcommand{\ofs}{6.7}
\newcommand{\drawlayer}[5][]{%
    \foreach \x/\n in {#5}{
        \node[#1] (\n) at (\x*#3+#4,-#2*\vsp) {\n};
    };%
}
\newcommand{\drawtwolayer}[4]{%
    \foreach \x/\n/\d in {#4}{
      \node[align=center] (\n) at (\x*#2+#3,-#1*\vsp) {\n\\(\scalebox{.7}\d)};
    };%
}
\newcommand{\drawedges}[3][]{%
    \foreach \x in {#3}{
        \draw[-stealth,thick] (#2) edge[#1] (\x);
    };
}
\newcommand{\drawalledges}[1]{%
    \foreach \x/\y in {#1}{
        \draw[-stealth,thick] (\x) edge (\y);
    }
}

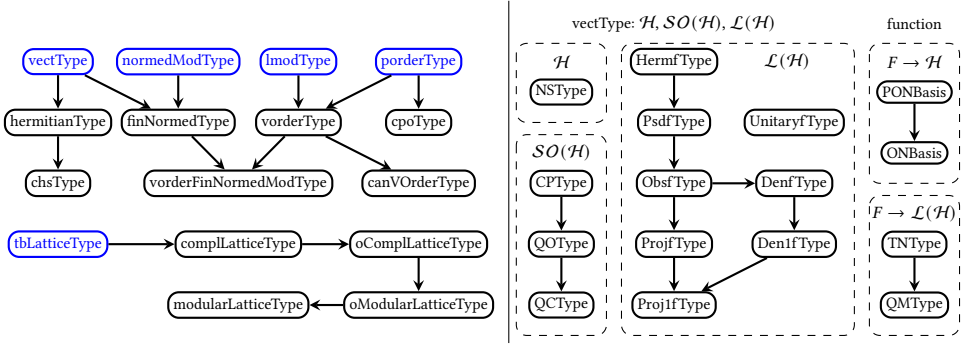
\begin{figure}[t]
\tiny
\begin{tikzpicture}[
    every node/.style={draw, thick, rectangle, minimum size=.2cm, inner sep=2pt,rounded corners=4pt},
    every child/.style={thick},sibling distance=1cm,
    edge from parent/.style={}]
    \drawlayer[color=blue]{1}{1.6}{0}{0/vectType,1/normedModType,2/lmodType,3/porderType};
    \drawlayer{2}{1.6}{0}{0/hermitianType,1/finNormedType,2/vorderType,3/cpoType};
    \drawlayer{3}{2.4}{0}{0/chsType,1/vorderFinNormedModType,2/canVOrderType};
    \drawedges{vectType}{hermitianType,finNormedType};
    \drawedges{normedModType}{finNormedType};
    \drawedges{lmodType}{vorderType};
    \drawedges{porderType}{vorderType,cpoType};
    \drawedges{hermitianType}{chsType};
    \drawedges{finNormedType}{vorderFinNormedModType};
    \drawedges{vorderType}{vorderFinNormedModType,canVOrderType};

    \drawlayer[color=blue]{4}{2.4}{0}{0/tbLatticeType};
    \drawlayer{4}{2.4}{0}{1/complLatticeType,2/oComplLatticeType};
    \drawlayer{5}{2.4}{0}{2/oModularLatticeType,1/modularLatticeType};
    \drawalledges{tbLatticeType/complLatticeType,complLatticeType/oComplLatticeType,oComplLatticeType/oModularLatticeType,oModularLatticeType/modularLatticeType};

    \draw (6,0) -- (6,-4.5);

    \node[draw=none] (pvectType) at (\ofs+1.5, -.3) {vectType: $\mathcal{H}, \mathcal{SO}(\mathcal{H}),\mathcal{L}(\mathcal{H})$};
    \draw[dashed,rounded corners=4pt] (\ofs-.6, -\vsp*.7) rectangle (\ofs+.6, -\vsp*2);

    \node[draw=none] (H) at (\ofs, -\vsp) {$\mathcal{H}$};
    \drawlayer{1.5}{3}{\ofs}{0/NSType};

    \node[draw=none] (LH) at (\ofs+3, -\vsp*1) {$\mathcal{L}(\mathcal{H})$};
    \drawlayer{1}{1.6}{\ofs+1.5}{0/HermfType};
    \drawlayer{2}{1.6}{\ofs+1.5}{0/PsdfType,1/UnitaryfType};
    \drawlayer{3}{1.6}{\ofs+1.5}{1/DenfType,0/ObsfType};
    \drawlayer{4}{1.6}{\ofs+1.5}{1/Den1fType,0/ProjfType};
    \drawlayer{5}{1.6}{\ofs+1.5}{0/Proj1fType};
    \drawedges{HermfType}{PsdfType};
    \drawedges{PsdfType}{ObsfType};
    \drawedges{ObsfType}{DenfType,ProjfType};
    \drawedges{DenfType}{Den1fType};
    \drawedges{ProjfType}{Proj1fType};
    \drawedges{Den1fType}{Proj1fType};

    \draw[dashed,rounded corners=4pt] (\ofs+.8, -\vsp*.7) rectangle (\ofs+3.9, -\vsp*5.5);

    \node[draw=none] (SO) at (\ofs, -\vsp*2.5) {$\mathcal{SO}(\mathcal{H})$};
    \drawlayer{3}{1.7}{\ofs}{0/CPType};
    \drawlayer{4}{1.7}{\ofs}{0/QOType};
    \drawlayer{5}{1.7}{\ofs}{0/QCType};
    \drawedges{CPType}{QOType};
    \drawedges{QOType}{QCType};
    \draw[dashed,rounded corners=4pt] (\ofs-.6, -\vsp*2.2) rectangle (\ofs+.6, -\vsp*5.5);


    \node[draw=none] (function) at (\ofs+4.7, -.3) {function};

    \node[draw=none] (FH) at (\ofs+4.7, -\vsp) {$F\to\mathcal{H}$};
    \drawlayer{1.5}{2}{\ofs+4.7}{0/PONBasis};
    \drawlayer{2.5}{2}{\ofs+4.7}{0/ONBasis};
    \drawedges{PONBasis}{ONBasis};
    \draw[dashed,rounded corners=4pt] (\ofs+4.1, -\vsp*.7) rectangle (\ofs+5.3, -\vsp*3);

    \node[draw=none] (FLH) at (\ofs+4.7, -\vsp*3.5) {$F\to\mathcal{L}(\mathcal{H})$};
    \drawlayer{4}{2}{\ofs+4.7}{0/TNType};
    \drawlayer{5}{2}{\ofs+4.7}{0/QMType};
    \drawedges{TNType}{QMType};
    \draw[dashed,rounded corners=4pt] (\ofs+4.1, -\vsp*3.2) rectangle (\ofs+5.3, -\vsp*5.5);
\end{tikzpicture}
  \caption{Hierarchy graph (left) and structures for quantum framework (right). The types colored by blue are types in mathcomp and mathcomp analysis.}
  \label{fig:hierachy_graph}
\end{figure}

\subsection{Formalization of \textbf{qwhile} and program logic}
\label{sec-concrete-syntax}

The abstract syntax \textbf{qwhile} language is deeply embedded in Coq, i.e., implemented as an inductive type \coqe{cmd}. The concrete syntax that support (abstract) data type, however, is implemented as a shallow embedding, i.e., as an instance of the abstract syntax, and thus offers the option to extend it in the future without a structural change of current development. 

\emph{Utility for data type.} CoqQ provides a rich basic constructs for typed states and unitary operators for both qubits and abstract types, 
including common 1/2-qubit gates, multiplexer, quantum Fourier bases/transformation, (phase) oracle (i.e., quantum access to a classical function) etc. CoqQ also allows to build unitary operator from partial information; for example, we use \coqe{VUnitary} to build the unitary $H_n$ which maps default state to uniform superposition state.

\emph{Composition of quantum variables.}
The composition of quantum variables is implemented in two level:
1). tensor of data type: for data type $\mT_1$ and $\mT_2$, notice that $\cH_{\mT_1}\otimes\cH_{\mT_2}\simeq\cH_{\mT_1\ast\mT_2}$, we define the tensor product of state $\bu_1\otimes\bu_2\in\cH_{\mT_1\ast\mT_2}$ and operators $A_1\otimes A_2\in\cL(\cH_{\mT_1\ast\mT_2})$;
2). choosing proper encoding maps: for two quantum variables \coqe{x$_1$ : vars T$_1$} and \coqe{x$_2$ : vars T$_2$} with disjoint domains, we carefully set the encoding map of \coqm{[x$_1$,x$_2$]} (which is of type \coqe{T$_1$*T$_2$}) for which the following properties holds:
$$
\hspace{1cm}
|(\mt_1,\mt_2)\>_{\overline{\mx}} = |\mt_1\>_{\mx_1}|\mt_2\>_{\mx_2},\ \ 
|\bu_1\otimes \bu_2\>_{\overline{\mx}} = |u_1\>_{\mx_1}|u_2\>_{\mx_2},\ \ 
(A_1\otimes A_2)[\overline{\mx}] = A_1[\mx_1]\otimes A_2[\mx_2].
$$
for all $\mt_i\in\mT_i, \bu_i\in\cH_{\mT_i}$ and $A_i\in\cL(\cH_{\mT_i})$ for $i = 1,2$.

\emph{Parameterized judgments.}
To avoid unnecessary duplication of inference rules, judgments are organized in a single definition, with two boolean parameters $\rm{pt} \in \{{\rm p}, {\rm t}\}$ and $\rm{st} \in \{{\rm s},{\rm n}\}$ which indicates the total (${\rm t}$) or partial (${\rm p}$) correctness and whether the judgment is saturated (${\rm s}$) or not (${\rm n}$) (saturated means that the inequality in Eqn. (\ref{eqn-total-local}) or (\ref{eqn-partial-local}) is replaced by ``=''). Most rules are formalized with parameter ${\rm pt}$ and/or ${\rm st}$, saying that they are sound for both partial/total correctness and/or saturated or not.

\emph{Support for for loops}. 
The proof rules for \textbf{for} loops rule make a full use of big operator library from mathcomp. These rules offer one-step derivation for \textbf{for} loops and provide a concise proofs for several circuit-building programs such as HLF algorithm, without the need for mathematical induction which is hard to use in practice if the index is a dependent type.

\subsection{Statistics}
Our Coq/SSReflect development is over 33,000 lines of code, evenly
split between definitions and proofs. Our new mathematical libraries
form the overwhelming majority of the development with 22000 lines of code in total. The other main
parts of the development are the labelled Dirac notation, the qwhile
language together with its utility, Hoare logic, and case studies, which respectively account
for about 2800, 5000, 1800 and 1300 lines of code. A detailed view
of the code statistics is given in Table~\ref{tab:number_of_lines}.

\begin{table}[htpb]
  \centering
  \caption{Code Metric.}
  \label{tab:number_of_lines}
  \begin{tabular}{|l|r|r|r|c|}
    \hline
     & \textbf{Spec.} & \textbf{Proof} & \textbf{Com.} & \textbf{Related files} \\
    \hline
    Complete partial order & 188 & 90 & 11 & cpo\\
    \hline
    Tactic for finite set  & 899 & 724 & 166 & setdec\\
    \hline
    Matrix norm / topology & 2744 & 3703 & 233 & mxpred mxnorm mxtopology \\
    \hline
    Tensor \& Hilbert spaces & 1017 & 1345 & 128 & xvector hermitian prodvect tensor \\
    \hline
    Linear maps / Super-operators & 3685 & 3590 & 237 & lfundef quantum \\
    \hline
    Subspace theory & 1466 & 1066 & 80 & orthomodular hspace \\
    \hline
    Labelled Dirac notation & 1562 & 1237 & 87 & dirac\\
    \hline
    Hilbert spaces over finite type & 1181 & 1466 & 125 & inhabited qtype\\
    \hline
    QWhile / quantum variables & 1245 & 790 & 176 & qwhile\\
    \hline
    Quantum Hoare logic & 917 & 919 & 87 & qhl \\
    \hline
    Case studies & 534 & 758 & 50 & example\\
    \hline
    Others & 329 & 360 & 3 & mcextra\\
    \hhline{|=|=|=|=|=|}
    \textbf{Total} & 15767 & 16048 & 1383 & \\
    \hline 
  \end{tabular}
\end{table}

\section{Case studies}\label{sec:examples}
We use CoqQ to prove the correctness of several well-known quantum algorithms from the literature; the code of the examples is given in Figure~\ref{fig-example-code}. Due to space limitations, we develop two examples in more detail, and only provide a short summary for the other examples.

\subsection{HSP algorithm}\label{sec:hsp}
Fig. \ref{fig-hsp-proof} outlines the main proof steps of the HSP algorithm. The proof is based on the assumption that the group $G$ is abelian, and uses three main facts from group theory:
\begin{itemize}
\item finite Abelian group $G$ is isomorphic to a Cartesian product (or direct sum, if one is familiar with group representation theory) of cyclic groups~\cite{Ser77}. Without loss of generality our formalization assumes that $G \triangleq \Pi_{0 \leq i < k} \mathbb{Z}_{p_i}$;
\item for every finite abelian group $G$ and subgroup $H$, $G$ is isomorphic to $H\times
 G/H$, where $G/H$ denote the quotient group. Elements of $G/H$ are subsets of $G$ called
 cosets, i.e.\, $J\in G/H$ iff there exists $g\in G$ such that $J = \{g + h\ |\ h\in H\}$. 
 Any coset $J$ has the same cardinality (number of elements) as $H$, i.e., $|J| = |H|$, and thus is always non-empty. We can arbitrarily choose an element from $J$, denoted by $({\rm repr}\ J)$, and $J = \{g + ({\rm repr}\ J)\ |\ g\in H\}$. Cosets are disjoint and the union of all cosets form $G$, which leads to
\begin{equation}
\label{eqn-hsp1}
    \sum_{g\in G}F(g) = \sum_{J : \text{coset of\ } H}\sum_{g\in H}F(g+(\text{repr\ }J))
\end{equation}
for arbitrary function $F : G\rightarrow T$ if $T$ is an additive abelian type.

\item The character function $\chi_g(h) \triangleq \prod_{m=0}^{k-1}e^{2\pi \imath g_m h_m/p_m}$ of $G$ where $g,h\in G$, satisfies:
\begin{equation}
\label{eqn-hsp2}
\chi_g(h) = \chi_h(g),\quad \chi_g(h_1+h_2)=\chi_g(h_1)\chi_g(h_2),\quad \sum_{h\in H}\chi_g(h) = \left\{
 \begin{array}{ll}
|H|&\forall h\in H. \chi_g(h) = 1\\
0&\text{otherwise}
\end{array}
\right..
\end{equation}
\end{itemize}
The proof interleaves applications of proof rules for program constructs, and calculations on labelled Dirac notation. The latter are denoted by $\Leftrightarrow$ in the figure, and are annotated with group theory when the calculation relies on the aforementioned facts from group theory. The final step of the proof applies the rule (R.Inner) to derive Eqn. (\ref{eqn-triple-HSP-in}) and (\ref{eqn-triple-HSP-notin})  from the current post-condition. The use of the rule (R.Inner) is very convenient here. Indeed, backward reasoning of HSP is relatively difficult since such derivation ignores the algebraic structure of the states during the execution that started from $|0\>_{\overline{\mx}}|\my_0\>_\my$. All proofs in the literature are forward, i.e., they show that the output is in a certain state.

\begin{figure}
\centering
\begin{align*}
&\textstyle \cg{1}\hspace{7cm}\text{Extra definitions/lemmas}\\
&\bullet\ \coqm{\textbf{for} $0\leq i < k$ \textbf{do} x$_i$ := $|0\>$;}\ ({\color{blue}\rm Ax.InFP'})
\hspace{0.4cm}\textstyle U[f] \triangleq \sum_{g\in G}\sum_{t\in X}|(g,t+f(g))\>\<(g,t)|\\
&\textstyle \cg{\bigotimes_{i=0}^{k-1}|0\>_{\mx_i}}\Longleftrightarrow\cg{|0\>_{\overline{\mx}}}
\hspace{3.4cm}\textstyle F_G \triangleq \frac{1}{\sqrt{|G|}}\sum_{g,h\in G}\chi_g(h)|g\>\<h|\\
&\bullet\ \coqm{y := $|$t$_0\>$ ;}\ ({\color{blue}\rm Ax.InF'})
\hspace{4.2cm}\textstyle\bigotimes_{i=0}^{k-1}\coqm{QFT}[\mx_i] = F_G[\overline{\mx}]
\\
&\textstyle \cg{|0\>_{\overline{\mx}}\otimes|\my_0\>_\my}
\\
&\bullet\ \coqm{\textbf{for} $0\leq i < k$ \textbf{do} x$_i$ := QFT[x$_i$];}\ ({\color{blue}\rm Ax.UTFP'})\\
&\textstyle \cg{\Big(\textstyle\bigotimes_{i=0}^{k-1}\coqm{QFT}[\mx_i]\Big)|0\>_{\overline{\mx}}\otimes|\my_0\>_\my}
\Longleftrightarrow\cg{F_G[\mx]|0\>_{\overline{\mx}}\otimes|\my_0\>_\my} \Longleftrightarrow\\
&\textstyle \cg{\frac{1}{\sqrt{|G|}}\sum_g|g\>_{\overline{\mx}}\otimes|\my_0\>_\my}\Longleftrightarrow
\cg{\frac{1}{\sqrt{|G|}}\sum_g|(g,\my_0)\>_{[\overline{\mx},\my]}}\\
&\bullet\ \coqm{[$\overline{\mx}$, y] := $U[f]$[$\overline{\mx}$, y];}\ ({\color{blue}\rm Ax.UTF'})
\\
&\textstyle \cg{U[f][\overline{\mx},\my]\frac{1}{\sqrt{|G|}}\sum_g|(g,\my_0)\>_{[\overline{\mx},\my]}} \Longleftrightarrow \cg{\frac{1}{\sqrt{|G|}}\sum_g|(g,\my_0+f(g))\>_{[\overline{\mx},\my]}}\xLongleftrightarrow[\text{\color{red}theory}]{\text{\color{red}group}}\\
&\cg{\frac{1}{\sqrt{|G|}}\sum_{J : \text{coset of\ } H}\sum_{g\in H}|g+(\text{repr\ }J)\>_{\overline{\mx}}|\my_0+f(\text{repr\ }J)\>_\my}
\\
&\bullet\ \coqm{\textbf{for} $0\leq i < k$ \textbf{do} x$_i$ := QFT[x$_i$].}\ ({\color{blue}\rm Ax.UTFP'})
\\
&\cg{\left(\bigotimes_{i=0}^{k-1}\coqm{QFT}[\mx_i]\right)\frac{1}{\sqrt{|G|}}\sum_{J : \text{coset of\ } H}\sum_{g\in H}|g+(\text{repr\ }J)\>_{\overline{\mx}}|\my_0+f(\text{repr\ }J)\>_\my}
\xLongleftrightarrow[\text{\color{red}theory}]{\text{\color{red}group}}\\
&\cg{\frac{1}{|H^\bot|}\sum_{g\in H^\bot}|g\>_{\overline{\mx}}\bigg[\sum_{J : \text{coset of\ } H}\chi_g(\text{repr\ }J)|\my_0+f(\text{repr\ }J)\>_\my\bigg]  }\\[0.2cm]
\hline\\[-0.4cm]
&({\color{blue}\rm R.Inner})\quad\forall\, g\in H^\bot,\ \models_{\rm t}^{\rm s} \left\{\frac{1}{|H^\bot|}\right\}\ HSP\ \left\{|g\>_{\overline{\mx}}\<g|\right\}\\
&({\color{blue}\rm R.Inner})\quad\forall\, g\notin H^\bot,\ \models_{\rm t}^{\rm s} \left\{0\right\}\ HSP\ \left\{|g\>_{\overline{\mx}}\<g|\right\}
\end{align*}
\caption{Proof outline for HSP algorithm. The predicate inside purple curly braces stands for $\cg{A}\triangleq\{AA^\dag\}$. All left-right arrows $\Leftrightarrow$ are the rewrites of the predicates.
CoqQ provide the built-in function \coqm{Oracle $f$} to construct $U[f]$ directly.
}
\label{fig-hsp-proof}
\end{figure}

\textit{Proof outline and interpretation of judgments.} We label the inference rules by blue color and summarize the main lemmas we used from group theory by red color in Fig. \ref{fig-hsp-proof}. 
We first derive the correct formula via forward reasoning:
\begin{equation*}
\cgrule{\rm t}{\rm s}{1}{HSP}{\frac{1}{|H^\bot|}\sum_{g\in H^\bot}|g\>_{\overline{\mx}}\bigg[\sum_{J : \text{coset of\ } H}\chi_g(\text{repr\ }J)|\mt_0+f(\text{repr\ }J)\>_\my\bigg]}.
\end{equation*}
and then finish the proof of Eqn. (\ref{eqn-triple-HSP-in}) and (\ref{eqn-triple-HSP-notin}), or more specifically, 
\begin{align*}
    & \forall\, g\in H^\perp,\ \models_{\rm t}^{\rm s} \left\{\frac{1}{|H^\perp|}\right\}\ HSP\ \left\{|g\>_{\overline{\mx}}\<g|\right\},\qquad
     \forall\, g\notin H^\perp,\ \models_{\rm t}^{\rm s} \left\{0\right\}\ HSP\ \left\{|g\>_{\overline{\mx}}\<g|\right\}.
\end{align*}
by employing rule (R.Inner). The superscript ``${\rm s}$'' means that both correctness are saturated, i.e., the pre and post-expectation are the same; the preconditions are scalars and thus interpreted as probability; the expectation of postcondition $|g\>_{\overline{\mx}}\<g|$ is the probability that we obtain $g$ if we measure $\overline{\mx}$ on computational basis.
In summary, these two Hoare triples exactly tell us that: after executing the quantum part of HSP algorithm and measuring the register $\overline{\mx}$, the  probability to obtain outcome $g$ is $\frac{1}{|H^\bot|}$ if $g\in H^\bot$ and $0$ if $g\notin H^\bot$, as we desired.

\emph{Statistics}. The full formalization is about 390 lines of code. It consists of 50+ lines for proving Eqn. (\ref{eqn-hsp1}) for general finite abelian groups, 60 lines for proving Eqn. (\ref{eqn-hsp2}), 10+ lines to set up variables, hypotheses and HSP algorithm and 250 lines for proving the correctness formulas Eqn. (\ref{eqn-triple-HSP-in}) and (\ref{eqn-triple-HSP-notin}).

\subsection{HHL algorithm}
The Harrow-Hassidim-Lloyd or HHL algorithm~\cite{HHL09}is a well-known quantum algorithm for solving linear systems of equations, i.e., finding a vector ${\bf x}$ such that $A{\bf x} = {\bf b}$ for given matrix $A$ and vector ${\bf b}$. \cite{ZYY19}  gives a pen-and-paper proof of the algorithm in Quantum Hoare Logic. We use CoqQ to provide a formalized proof.

For simplicity, we assume $A$ is a full-rank Hermitian operator in $\cH_{\mathbb{Z}_m}$ (i.e., A is of $m$-dimensional matrix) with diagonal decomposition $A = \sum_{j < m}\lambda_j|u_j\>\<u_j|$ where $\{|u_j\>\}$ is an orthonormal basis of $\cH_{\mathbb{Z}_m}$ and assume $|b\>\in\cH_{\mathbb{Z}_m}$ an normalized state (i.e., $\||b\>\|=1$). To make the algorithm exact, we further assume that there exists $t_0\in \mathbb{R}$, for all $0\le j\le m$, $\lambda_jt_0$ is multiple of $2\pi$, i.e., $\delta_j = \frac{\lambda_jt_0}{2\pi}\in \{1,2,\cdots,n-1\}.$ 

The algorithm operates on three quantum variables: 1. \coqm{q : vars $\mathbb{Z}_m$} which is used to store the input data $|b\>$ -- achieved by applying a unitary operator $U_b$ which transforms $|0\>$ into $|b\>$; 2. \coqm{p : vars $\mathbb{Z}_n$} which acts as the control system and is used to store the value of $\delta_j$; and 3. \coqm{r : vars $\mathbb{B}$} which is used as the guard of while loop -- indicating if the subroutine succeeds -- done by employing a controlled unitary $U_c : \cL(\cH_{\mathbb{Z}_n\ast\mathbb{B}})$ with some suitable parameter $C$ such that:
\begin{align*}
\textstyle
U_c|(0,0)\> = |(0,0)\>; \quad  U_c|(i,0)\> = |i\>\Big(\sqrt{1-\frac{C^2}{i^2}}|0\> + \frac{C}{i}|1\>\Big)\ \forall i = 1,2,\cdots,n-1.
\end{align*}
Writing the algorithm uses several built-in functions of CoqQ. For example, 
$U_b$ and $U_c$ are constructed by the built-in functions \coqe{PUnitary} and \coqe{VUnitary} which provide a unitary from the partial information (e.g., we only provide $|0\>\mapsto|b\>$ for $U_b$). The key transformation of HHL is the multiplexer of the function $f : k \mapsto e^{\imath Akt_0/n}$; that is,
$U_f\triangleq\coqm{Multiplexer}(f) = \sum_{k<n}|k\>\<k|\otimes e^{\imath Akt_0/n}$.

The HHL program together with a proof outline is displayed in Fig. \ref{fig-hhl-proof}.
It is easy to see that the solution for the linear equation $A|x\> = |b\>$ is 
$|x\> = c\sum_{j = 1}^N \frac{\beta_j}{\lambda_j}|u_j\>$ with $\beta_j\triangleq\<j|b\>$
up to some unimportant scale factor where $c$ is only used to normalize $|x\>$. The correctness of HHL algorithm can be formulated as:
\begin{equation}
\label{eqn-triple-HHL}
\cgrule{\rm p}{}{1}{HHL}{|x\>_\mq}
\end{equation}
Informally, whenever the program terminates, then the variable $\mq$ is in state $|x\>$.

The proof makes use of subspace theory based on projection representation -- which simplifies the the proof of L\"owner order property. Roughly speaking, a state $|v\>$ with norm $\||v\>\|\le 1$ lies in some subspace $V$ (convertible to the projection) must implies $|v\>\<v|\sqsubseteq V$.

\emph{Statistics.} The code in total is about 280 lines, including 90 lines for set up the parameters and simple properties of these parameters, 15 lines to define the HHL program and 170 lines for prove the Eqn. (\ref{eqn-triple-HHL}).

\begin{figure}
\centering
\begin{align*}
&\cg{1}\hspace{7cm}\text{Extra definitions/lemmas}\\
&\bullet\ \coqm{p := |0\>;\ q := |0\>;\ r := |0\>;}\ {\color{blue}\rm (Ax.InF')}
\hspace{0.56cm}P \triangleq |0\>_\mp\<0|\otimes I_q\otimes |0\>_\mr\<0|\\
&\cg{|0\>_\mp|0\>_\mq|0\>_\mr}\xLongrightarrow{{\color{blue}\rm (R.Or)}}\{R\}
\hspace{3.17cm}Q \triangleq |0\>_\mp\<0|\otimes |x\>_\mq\<x|\otimes|1\>_\mr\<1|\\
&\bullet\ \coqm{\textbf{while}\ {\rm meas}[r] = 0\ \textbf{do}}\ {\color{blue}\rm (R.LP.P)}
\hspace{2cm}R \triangleq P + Q\\
&\qquad\{P\}\xLongrightarrow{{\color{blue}\rm (R.TI)}}\cg{|0\>_\mp|0\>_\mr}
\hspace{3.1cm}\textstyle R = |0\>_\mr\<0|P|0\>_\mr\<0| + |1\>_\mr\<1|Q|1\>_\mr\<1|\\
&\qquad\bullet\ \coqm{q := |0\>;}\ {\color{blue}\rm (Ax.InF')}
\hspace{2.98cm}\textstyle|v_j\>_\mr \triangleq \sqrt{1-\frac{C^2}{\delta_j^2}}|0\>_\mr + \frac{C}{\delta_j}|1\>_\mr,\quad\forall 1\le j\le n\\
&\qquad\cg{|0\>_\mp|0\>_\mq|0\>_\mr}\\
&\qquad\bullet\ \coqm{q := $U_b$[q];}\ {\color{blue}\rm (Ax.UTF')}\\
&\qquad\cg{|0\>_\mp(U_b[\mq]|0\>_\mq)|0\>_\mr}\Leftrightarrow\cg{|0\>_\mp|b\>_\mq|0\>_\mr}\\
&\qquad\bullet\ \coqm{p := $H_u$[p];}\ {\color{blue}\rm (Ax.UTF')}\\
&\qquad\textstyle\cg{(H_u[\mp]|0\>_\mp)|b\>_\mq|0\>_\mr}\Leftrightarrow\cg{\frac{1}{\sqrt{n+1}}\sum_{\tau : [n+1]}|\tau\>_\mp|b\>_\mq|0\>_\mr}\\
&\qquad\bullet\ \coqm{[p,q] := $U_f$[p,q];}\ {\color{blue}\rm (Ax.UTF')}\\
&\qquad\textstyle\cg{\frac{1}{\sqrt{n+1}}\sum_\tau(U_f[\mp,\mq]|\tau\>_\mp|b\>_\mq)|0\>_\mr}\Leftrightarrow
\cg{\frac{1}{\sqrt{n+1}}\sum_j\big(\sum_\tau\beta_je^{\imath \tau\lambda_jt_0/(n+1)}|\tau\>_\mp\big)|u_j\>_\mq|0\>_\mr
}\\
&\qquad\bullet\ \coqm{p := IQFT[p];}\ {\color{blue}\rm (Ax.UTF')}\\
&\qquad\textstyle\cg{\frac{1}{\sqrt{n+1}}\sum_j\big(\coqm{IQFT}[\mp]\sum_\tau\beta_je^{\imath \tau\lambda_jt_0/(n+1)}|\tau\>_\mp\big)|u_j\>_\mq|0\>_\mr}
\Leftrightarrow\cg{\sum_j\beta_j|\delta_j\>_\mp|u_j\>_\mq|0\>_\mr}
\\
&\qquad\bullet\ \coqm{[p,r] := $U_c$[p,r];}\ {\color{blue}\rm (Ax.UTF')}\\
&\qquad\textstyle\cg{\sum_j\beta_j|u_j\>_\mq(U_c[\mp,\mr]|\delta_j\>_\mp|0\>_\mr)}\Leftrightarrow \cg{\sum_j\beta_j|\delta_j\>_\mp|u_j\>_\mq|v_j\>_\mr}\\  
&\qquad\bullet\ \coqm{p := QFT[p];}\ {\color{blue}\rm (Ax.UTF')}\\
&\qquad\textstyle\cg{\sum_j\beta_j(\coqm{QFT}[\mp]|\delta_j\>_\mp)|u_j\>_\mq|v_j\>_\mr}\Leftrightarrow
\cg{\frac{1}{\sqrt{n+1}}\sum_{j,\tau}\beta_je^{\imath 2\pi \delta_j\tau/T}|\tau\>_\mp|u_j\>_\mq|v_j\>_\mr}
\\
&\qquad\bullet\ \coqm{[p,q] := $U_f^\dag$[p,q];}\ {\color{blue}\rm (Ax.UTF')}\\
&\qquad\textstyle\cg{\frac{1}{\sqrt{n+1}}\sum_{j,\tau}\beta_je^{\imath 2\pi \delta_j\tau/T}(U_f^{-1}[\mp,\mq]|\tau\>_\mp|u_j\>_\mq)|v_j\>_\mr}\Leftrightarrow
\cg{\frac{1}{\sqrt{n+1}}\sum_\tau|\tau\>_\mp\sum_j\beta_j|u_j\>_\mq|v_j\>_\mr}
\\
&\qquad\bullet\ \coqm{p := $H_n^\dag$[p];}\ {\color{blue}\rm (Ax.UTF')}\\
&\qquad\textstyle\cg{\big(H_n^\dag[\mp]\frac{1}{\sqrt{n+1}}\sum_\tau|\tau\>_\mp\big)\sum_j\beta_j|u_j\>_\mq|v_j\>_\mr}
\Leftrightarrow\cg{\underline{|0\>_\mp\sum_j\beta_j|u_j\>_\mq|v_j\>_\mr}}
\\
&\qquad\textstyle\xLongrightarrow[{\color{red}\rm subspace\ theory}]{{\color{blue}\rm (R.Or)}}\{R\}
\textstyle\hspace{3cm}{\color{red}= \underline{C|0\>_\mp|x\>_\mq|1\>_\mr} + \underline{|0\>_\mp\Big(\sum_j\beta_j\sqrt{1-\frac{C^2}{\delta_j^2}}|u_j\>_\mq\Big)|0\>_\mr}}\\
&\bullet\ \coqm{\textbf{do}}\ {\color{blue}\rm (R.LP.P)}
\textstyle\hspace{5.5cm}{\color{red}\in Q \hspace{1.3cm}+\hspace{1.3cm} \in P}\\
&\{Q\}\xLongrightarrow{{\color{blue}\rm (R.Or)}}\cg{|x\>_\mq}
\end{align*}
\caption{Proof outline for HHL algorithm. Left-right arrow represents the rewrite of predicates. \coqm{QFT} and \coqm{IQFT} are built-in (inverse) quantum Fourier transformation for $\bZ_{p}$ type. $H_n$ is the built-in unitary that maps default state ($|0\>$ here) to the uniform superposition state. }
\label{fig-hhl-proof}
\end{figure}

\begin{figure*}
  \begin{minipage}{.49\textwidth}
\begin{coq}
(* x : vars bool *)
(* s : sequence of (vars bool)) *)
Definition QFT_sub x s := 
  \for i < size(s) \do 
    [x,s$_{\mi}$] := CU(Ph($e^{\pi\imath/2^{\mi+1}}))$[x,s$_{\mi}$].

Fixpoint QFT_iter s :=
  match s with
  | [::] => $\askip$
  | x :: t => x := H[x]; 
              (QFT_sub x t); 
              (QFT_iter t)
  end.
 
(* s : n-tuple of (vars bool) *)
Definition QFT_cir s :=
  QFT_iter s;
  rev_circuit s.

Definition QPE :=
  x := $|0\>$;
  x := $H_n$[x];
  [x,y] := Multiplexer(fun i => $U^\mi$)[x,y];
  x := IQFT[x].

\end{coq}
  \end{minipage} 
  \begin{minipage}{.49\textwidth}
\begin{coq}
(* x : n-tuple of (vars bool)) *)
Definition ParaHadamard := 
  \for i < n \do x$_\mi$ := H[x$_\mi$].

(* x : n-tuple of (vars T)) *)
Definition rev\_circuit := 
  \for i < $\lfloor$n/2$\rfloor$ \do 
    [x$_\mi$,x$_{\mn-\mi}$] := SWAP[x$_\mi$,x$_{\mn-\mi}$].

Definition HLF := 
  \for i \do x$_\mi$ := $|0\>$;
  \for i \do x$_\mi$ := H[x$_\mi$];
  \for i $\in$ SD \do x$_\mi$ := S[x$_\mi$];
  \for i $\in$ SS \do 
    [x$_{\mi_1}$,x$_{\mi_2}$]:= CZ[x$_{\mi_1}$,x$_{\mi_2}$];
  \for i \do x$_\mi$ := H[x$_\mi$].

Definition Grover (r : nat) :=
  x := |t$_0\>$;
  x := $H_n$[x];
  \for i < r \do (
    x := PhOracle(f)[x];
    x := $H_n^\dag$[x];
    x := PhOracle(fun i => i == t$_0$)[x];
    x := $H_n$[x];).
\end{coq}
  \end{minipage}
\caption{Programs for Parallel Hadamard, reverse circuit, QFT circuit, Hidden linear function, Quantum phase estimation and Grover search algorithm. \coqm{CU, Ph, H, CZ} are the built-in controlled-unitary gate, parameterized phase gate, Hadamard gate and controlled-Pauli Z gate respectively.
\coqm{Multiplexer, SWAP, PhOracle, $H_n$} are built-in multiplexer, SWAP gate, phase oracle and uniform
transformation (from default state ($|0\>$ in \coqm{QPE} and \coqm{$|$t$_0\>$} in \coqm{Grover}) to uniform
superposition state). The \coqm{QFT\_cir} is a parameterized circuit and thus written using a fixpoint definition.
}
\label{fig-example-code}
\end{figure*}

\subsection{Parallel Hadamard}
Parallel Hadamard is a circuit that converts the initial state of a quantum circuit to a uniform superposition state. Parallel Hadamard is a key step to leverage the power of quantum computation and is frequently used at the beginning or the end of a circuit implementation of algorithms. The program is quite simple, as shown in Fig. \ref{fig-example-code}. We prove:
\begin{align}
&\textstyle\cgrule{\rm pt}{\rm st}{\bigotimes_i|0\>_{\mx_i}}{\coqm{ParaHadamard}}{\bigotimes_i|+\>_{\mx_i}}\quad
\cgrule{\rm pt}{\rm st}{\bigotimes_i|+\>_{\mx_i}}{\coqm{ParaHadamard}}{\bigotimes_i|0\>_{\mx_i}}
\label{eqn-parahadamard1}
\\
&\textstyle\cgrule{\rm pt}{\rm st}{|b\>_{\overline{\mx}}}{\coqm{ParaHadamard}}{\frac{1}{\sqrt{2^n}}\sum_t(-1)^{\sum_ib_it_i}|t\>_{\overline{\mx}}}
\label{eqn-parahadamard2}
\end{align}
The \coqe{ParaHadamard} transforms $\bigotimes_i|0\>_{\mx_i}$ to $\bigotimes_i|+\>_{\mx_i}$ and vice versa as we expected, or more generally, transforms $|b\>_{\overline{\mx}}$ to $\frac{1}{\sqrt{2^n}}\sum_t(-1)^{\sum_ib_it_i}|t\>_{\overline{\mx}}$.

\emph{Statistics.} We use (Ax.UTPF') to reasoning about Eqn. (\ref{eqn-parahadamard1}) in one step with only 3 lines of proof code each and Eqn. (\ref{eqn-parahadamard1}) with about 20 lines since we should derive $\sum_ib_it_i$; it takes in total 60 lines for proving several goals for tuple and finite functions.

\subsection{QFT circuit and reverse circuit}
As one of the most fundamental building blocks in designing quantum algorithm, verifying the correctness of QFT circuit is a common task for a program verifier. The QFT circuit is parameterized by a meta variable $n$ (its size) and thus is defined using fixpoint function rather than the concrete syntax. It also employs a reverse circuit at the end, which is used to reverse the order of qubits.  We show:
\begin{align}
&\cgrule{\rm pt}{\rm st}{|t\>_{\overline{\mx}}}{\coqm{rev\_circuit}}{|t\>_{\overline{{\tt r\tt e\tt v(\tt x)}}}}\quad
\cgrule{\rm pt}{\rm st}{|t\>_{\overline{{\tt r\tt e\tt v(\tt x)}}}}{\coqm{rev\_circuit}}{|t\>_{\overline{\mx}}}
\label{eqn-rev-cir}
\\
&\textstyle
\cgrule{\rm pt}{\rm st}{|b\>_{\overline{{\tt s}}}}{\coqm{QFT\_cir}}{|\coqm{QFTbv } b\>_{\overline{{\tt s}}}}\quad \text{where\ }
|\coqm{QFTbv } b\> = \frac{1}{\sqrt{2^n}}\sum_{t:\{0,1\}^n}e^{2\pi\imath f(b)f(t)/2^n}|t\>
\label{eqn-qft-cir}
\end{align}
where the $f$ converts a bit string ($\mathbb{B}$ tuple) to a natural number $\mathbb{N}$.
The reverse circuit \coqe{rev\_circuit} simply reverses the order of a tuple of variables $\overline{\mx}$ (with arbitrary types rather than just qubit).
Eqn. \ref{eqn-qft-cir} can be easily interpreted as converting $|b\>_{\overline{\coqm{s}}}$ to its corresponding QFT basis $|\coqm{QFTbv } b\>$. 

\emph{Statistics for reverse circuit.}  It is about 80 code to formalize reverse circuit and finish the proof of 
Eqn. (\ref{eqn-rev-cir}); it is slight longer since we need to show the side-condition of using (Ax.UTFP) -- disjointness of each SWAP.

\emph{Statistics for QFT circuit.} The formalization and proof uses mathematical induction and take about 140+ lines of codes in total (excluding the code of the reverse circuit).

\subsection{BGK algorithm}
The Bravyi-Gosset-Konig, or BGK, algorithm~\cite{BGK18} is an algorithm to solve the hidden linear function (HLF). Suppose $A$ is a $n\times n$ symmetric Boolean matrix, the goal of HLF problem is to find a Boolean vector $z\in\{0,1\}^n$ such that 
$$\textstyle\forall x, (Ax = 0\mod 2)\rightarrow \big(q(x) = 2\sum_iz_ix_i\mod 4\big)$$
where $q(x)\triangleq\sum_{i,j}A_{ij}x_ix_j\mod 4$.
We define two set ${\tt S\tt D} = \{i\ |\ A_{ii} = 1\}$ and ${\tt S\tt S} = \{(i,j)\ |\ i < j \text{ and } A_{ij} = 1\}$ to specify the code.
As summarized in Eqn. (4) in \cite{BGK18}, we verify:
\begin{align}
\textstyle
\cgrule{\rm pt}{\rm st}{1}{\coqm{HLF}}{\frac{1}{2^n}\sum_{z:\{0,1\}^n}
\Big(\sum_{k:\{0,1\}^n}\imath^{q(k)+2\sum_ik_iz_i}\Big)|z\>_{\overline{\mx}}}.
\label{eqn-BGK}
\end{align}

The algorithm is implemented as a circuit-building program that working on a 2-dimensional grids of qubits.

\emph{Statistics.} All code is about 160 lines: 30+ lines to set up the variables, parameters and algorithm; and 120+ lines to finish the proof of Eqn. (\ref{eqn-BGK}).

\subsection{QPE}
Quantum phase estimation (QPE) is a quantum algorithm that is often used as a subroutine in other quantum algorithms to estimate the phase (eigenvalue) of an eigenvector of a unitary operator.
Given a unitary operator $U$ and an eigenvector $|\phi\>$ of $U$; the goal is to find an approximation $0\le\theta<1$ such that $U|\phi\> = e^{2\pi\imath\theta}$. The correctness of QPE is stated as:
\begin{align}
\textstyle
\forall a < n,\ \cgrule{\rm pt}{\rm st}{c(a)|\phi\>_\my}{\coqm{QPE}}{|a\>_\mx|\phi\>_\my}\quad\text{with\ }
c(a) \triangleq\sum_{j<n}e^{2\pi\imath(a/n-\theta)j/n}.
\label{eqn-qpe}
\end{align}
where \coqe{y : vars T} stores $|\phi\>$ and \coqe{x : vars $\bZ_n$} is the control system that is used to approximate $n\theta$.

This judgment tells us if we measure register $\mx$ at the end, we have probability $|c(a)|^2$ to obtain outcome $a$. A straightforward calculation shows that, if $n\theta\in\mathbb{N}$, then we have probability 1 to get $n\theta$; otherwise, we have at least probability $4/\pi^2$ to obtain ${\rm round}(n\theta)$ (i.e., the closest integer to $n\theta$).

\emph{Statistics.} 
Setting up algorithms and proving Eqn. (\ref{eqn-qpe}) only takes 40+ lines of code; we take another 50 lines to show the property of function $c(a)$, i.e., $|c({\rm round}(n\theta))|$ is 1 if the algorithm is exact, and at least $2/\pi$ otherwise. 

\subsection{Grover's algorithm}

Grover's algorithm is a quantum algorithm for unstructured data search and offers a quadratic speedup compared to the classical algorithm.
We assume that the type of data is $\mT$ and the given function $f : \mT \rightarrow \mathbb{B}$ which can be accessed by a phase oracle (i.e., modeled as \coqe{PhOracle[$f$]}). Let 
$|v\>\triangleq\frac{1}{\sqrt{|f|}}\sum_{i : f(i)=1}|i\>$ the superposition of all solutions (i.e., $f(i) = 1$). We show that if we run the subroutine for $r$ times, the following Hoare triple holds:
\begin{align}
\textstyle
    \models_{\rm pt}\left\{\sin^2((2r+1)t)\right\}\coqm{Grover(r)}\left\{\sum_{i : f(i)=1}|i\>_\mx\<i|\right\}
    \quad \text{with\ }t\triangleq\arcsin{\sqrt{\frac{|f|}{|T|}}}.
\end{align}
It implies that if we measure $\mx$ at the end, we have at least the probability $\sin^2((2r+1)t)$ to obtain a solution of $f$. With a proper choice of $r$, i.e., let $\sin^2((2r+1)t)$ close to 1, the Grover's algorithm succeed with high probability.

\emph{Statistics.} 
The code is about 180 lines, including 30+ lines to set up the parameters and algorithms and 140+ lines for proving the Hoare triple.

\section{Related work}\label{sec:related}
There is a large body of work in the design, implementation, and verification of quantum programs. For convenience, we distinguish approaches that are supported
by artefacts (formalizations in proof assistants, verification tools) and theoretical approaches. For space reasons, we exclude approaches based on testing and program analyses. 

\subsubsection*{Mechanized approaches}
Table~\ref{tab:related-work} summarizes the main differences between CoqQ and some other verification tools for quantum programs, based on the four criteria discussed in the introduction. We comment on the tools below.

QWIRE~\cite{paykin2017qwire,RPZ18} is a Coq formalization of quantum programs written in a circuit-like language. The formalization includes a denotational semantics of programs in terms of density matrices, and has been used to verify several interesting algorithms. A recent extension of QWIRE, called ReQWIRE~\cite{rand2019reqwire} develops a verified compiler to compile classical circuits to reversible quantum circuits. QWIRE does not include a program logic. QWIRE is built on top of the standard libary of Coq for the theory of real numbers, and builds its own libraries for complex number and matrix theory. Interestingly, the authors of QWIRE report that mathcomp was also considered as external library for early development of QWIRE, but due to the overhead caused by dependent types, the authors use phantom type \cite{RPZ18a} instead.  In contrast, the use of mathcomp is more important in our setting, as we aim to support general notions of state. 

SQIR~\cite{HRH21,hietala2020proving} is a Coq formalization for formal verification of quantum programs. SQIR provides a semantics of programs based on a density matrix representation of quantum states, and proves the correctness of a quantum circuit optimizer. The formalization does not include a program logic. Like QWIRE, SQIR is built on top of the standard libary of Coq. For the latter work that extracts OpenQASM programs from the Coq representation of SQIR programs, the Coq extraction mechanism is in the Trusted Computing Base.

The QHLProver~\cite{liu2019formal} is an Isabelle formalization for formal verification of quantum programs based on quantum Hoare logic~\cite{Ying11}. Their formalization includes a denotational semantics of \textbf{qwhile} programs, and a Hoare logic that is proved sound with respect to the program semantics.
The prover is used to verify several examples, including Grover's algorithm. However, a main difference with our work is that their formalization is based on a matrix representation of states, rather than an abstract representation. This makes the formalization of examples such as HSP cumbersome. The formalization uses
the library JNF -- Jordan\_Norm\_Forms \cite{thiemann2016formalizing} for matrices, and the library DL -- Deep\_Learning \cite{bentkamp2019formal} for tensors.

qrhl-tool~\cite{Unr19} is an Isabelle formalization for formal verification of quantum programs. Programs in qrhl-tool are written in a high-level language that support rich types for quantum variables. Programs can be verified using quantum relational Hoare logic (QRHL)~\cite{Unr19}. The main application of the qrhl-tool is security proofs of (post-quantum) cryptographic constructions. Earlier versions of qrhl are not foundational: program semantics are not built from first principles, and the program logic is not proved sound with respect to the denotational semantics. Recently, Caballero and Unruh developed CBO -- Complex Bounded Operator~\cite{caballero2021complex}, an Isabelle library that is used for modelling assertions. qrhl-tool uses CBO and many other libraries.

Isabelle Marries Dirac~\cite{bordg2021certified} is an Isabelle formalization for formal verification of quantum programs. The formalization uses a shallow embedding of quantum circuits based on the JNF library for matrices~\cite{thiemann2016formalizing}. The formalization also provides a rudimentary encoding of Dirac notation; for instance, no big operators are supported. The formalization is used to verify several algorithms, and helped uncover a bug in a published proof of the Quantum Prisoner Dilemma.

Qbricks~\cite{chareton2020deductive} is a verification framework for circuit-building quantum programs. The framework is a classic automated verification framework that support automated verification of rich specifications. Concretely, QBricks targets SMT-solvers via the Why3 verification platform. The encoding of programs into SMT-clauses is based on the path-sum representation of quantum states introduced in~\cite{amy2018towards}, and uses a custom formalization of algebra and matrices starting from a few axiomatic definitions. The verification approach is proved sound on paper, and the framework is used to verify many of the algorithms that are verified using CoqQ.

\subsubsection*{Theoretical approaches}
\label{sec-related-work-other}
We compare our approach with Ying's work on quantum Hoare logic and recent work on quantum separation logic. Then we briefly discuss other approaches.

\paragraph*{Comparison with \cite{Ying11}}
\cite{Ying11} proposes the first sound and relatively complete proof system for \textbf{qwhile}-programs. Ying's seminal work was subsequently extended in many directions. \cite{Unr19a,FY21} extend the program logic to programs with quantum and classical variables. \cite{YZL18,YZL22} extend the logic to parallel programs. \cite{ZYY19} proposes a variant of quantum Hoare logic that use projectors as
pre- and postconditions.

The proof rules in these systems are similar to ours in many ways. The rules for basic constructs are inspired from \cite{Ying11}, the rules for \textbf{for} loops are similar to  \cite{YZL18,YZL22}, and many structural rules are inspired from \cite{Ying19}. One minor difference is that our rules use linear operators rather than observables. This minimizes
the number of side-conditions in proof rules. Note that the rule (R.Inner) is new.

\paragraph*{Comparison with Quantum Separation Logic}
Separating conjunction is a logical construct that is used by separation (and other resource-aware) logics to model disjointness between two systems. Separating conjunction was originally used to model spatial disjunction on heaps.  There have been two recent proposals~\cite{ZBH21,LLSS22} to use separating conjunction for capturing separability (vs. entanglement) of quantum states. However, these proof systems are not powerful enough for programs with highly entangled subroutines. It remains to be seen whether this is a limitation of these two proof systems, or a fundamental limitation of separating conjunction. From the perspective of users familiar with quantum physics, one potential drawback of using separation conjunction (rather than labels, which enforce separation syntactically) is that the connection with labelled Dirac notation is lost.

\paragraph*{Generalizations and other approaches}
Two recent works~\cite{BHY19,Unr19,LU21,BBF21} develop relational Hoare logics for quantum programs. A promising direction for future work is to enhance our formalization to support relational reasoning.

There are also many alternative approaches to verify quantum programs based on dynamic logic, temporal logic, Kleene algebra, type theory and process algebra, see for instance~\cite{brunet2004dynamic,akatov2005logic,baltag2006lqp,feng2007proof,kakutani2009logic,yu2019quantum, singhal2020quantum,peng2022algebraic}. A long term goal would be to build verified program verifiers for some of these  approaches.

Another approach is the ZX-calculus~\cite{coecke2011interacting} that was proposed for reasoning about linear maps between qubits. The ZX-calculus originates from earlier work on categorical foundations of quantum physics~\cite{coecke2006kindergarten, abramsky2009categorical} and has found several useful applications in simplification~\cite{duncan2020graph} and equivalence checking~\cite{peham2022equivalence} of quantum circuits. Recent work~\cite{jeandel2018diagrammatic,hadzihasanovic2018two,vilmart2019near} provides some complete axiomatizations of the ZX-calculus.
However, we are not aware of any mechanization of the calculus.

\subsubsection*{Language design}
The design of quantum programming languages is an active area of research. Many existing works emphasize principled foundations, and develop semantics and type systems/program analyses to guarantee programs are well-behaved. Due to space limitations, we only cite a few examples.  Quipper~\cite{green2013quipper}, Q\#~\cite{Qsharp} and Tower~\cite{yuan2022tower} allow users to build complex data types from qubit; Silq~\cite{bichsel2020silq} supports automatic uncomputation of quantum programs with the help of their elaborate type system; Twist~\cite{YMC22} develops a type system to help people manage entanglements in their quantum programs.

\section{Conclusion}
We have introduced CoqQ, the first verified quantum program verifier
for a high-level quantum programming language. One main strength of
CoqQ is to leverage state-of-the-art mathematical libraries, i.e.\,
mathcomp and mathcomp analysis. We have illustrated the benefits of
CoqQ by mechanizing several examples of the literature. However, there
remains many further steps to improve its expressiveness and its
usability. A first step is to introduce classical variables, so as to
support reasoning about programs that mix quantum and classic
computations. Another important step is to support for data
structures, leveraging a recent proposal~\cite{yuan2022tower} to
incorporate data structures in quantum programs. Naturally, it would
also be interesting to provide better support for automating recurring
mundane tasks. In the longer term, we would like to use CoqQ as the
basis for a verified quantum software toolchain. A key element would be to
develop a formally verified compiler from \textbf{qwhile} to quantum
circuits (akin to CompCert~\cite{compcert} for the Verified Software
Toolchain~\cite{vst}). Another exciting direction would be to build a
certified abstract interpreter, following~\cite{YP21} (akin to
Verasco~\cite{verasco} for CompCert).

\begin{acks}                            
We thank Cyril Cohen and Christian Doczkal for their valuable discussion.
\end{acks}

\bibliographystyle{ACM-Reference-Format}
\bibliography{main}

\newpage
\begin{appendix}
{\Large \textbf{Supplementary material}}

\section{Abstract Linear algebra}

We provide some basic definitions and remarks of the abstract linear algebra in this section that complement Section \ref{sec:prelim}. We write $\bF$ for a field, $\bC$ for the complex numbers. We write $\imath$ for the imaginary of $\bC$ and $\overline{c}$ stands for the conjugate of $c\in\bC$.

\begin{defn}[Vector space over $\bF$]
A vector space (also called a linear space) over field $\bF$ is a set $V$ equipped with addition $+ : V \times V \ra V$ (write as $\bu + \bv$ for $\bu,\bv\in V$) and scalar multiplication $\bF \times V \ra V$ (write as $a\bu$ for $a\in\bF$ and $\bu\in V$) with following properties:
\begin{enumerate}
    \item $(V,+)$ forms an abelian group;
    \item Associativity: $a(b\bv) = (ab)\bv$ for all $a,b\in\bF$ and $\bv\in V$;
    \item Distributivity: $a(\bu+\bv) = a\bu + a\bv$, $(a+b)\bv = a\bv + b\bv$ for all $a,b\in\bF$ and $\bv\in V$;
    \item Scalar multiplicative identity: $1\bv = \bv$ for all $\bv\in V$.
\end{enumerate}
\end{defn}
The field $\bF$ itself is a linear space with multiplication as scalar multiplication.

\begin{defn}[linear combination, span, linear independence and basis]
Let $\{\bu_i\}$ be a set of vectors in $V$ and $\{a_i\}$ a set of scalars in $\bF$ with index $i$ being finite or infinite. We call $\sum_i a_i\bu_i$ (if it converges when $i$ is infinite) the linear combination of $\bu_i$. 

Let $S\subseteq V$, its span $\spans\{\bu_i\}$ is the set of all finite linear combinations of (subset of) $S$. $\spans\{\bu_i\}$ is again a linear space.

We say $\{\bu_i\}$ is linearly dependent if there exists $\{a_i\}$ which is not all zero and $\sum_ia_i\bu_i=0$. $\{\bu_i\}$ is linearly independent if it is not linearly dependent.

We say $\{\bu_i\}$ is a basis of $V$ if $\{\bu_i\}$ is linearly independent and any $\bv\in V$ can be written as a linear combination of $\{\bu_i\}$. For any $V$, there exists a basis; and any basis of $V$ has the same cardinality -- called the dimension of $V$, write as $\dim(V)$. 
If $\dim(V)$ is finite, we call $V$ a finite-dimensional linear space. 
\end{defn}

\begin{defn}[Linear map]
Suppose $U$ and $V$ are two linear space over the same field $\bF$. We say a map $f : U \ra V$ is linear if it satisfies
\begin{enumerate}
    \item Additivity: $f(\bu+\bv) = f(\bu) + f(\bv)$ for all $\bu,\bv\in U$;
    \item Homogeneity: $f(au) = af(\bu)$ for all $a\in\bF$ and $\bu\in U$.
\end{enumerate}
\end{defn}

Given the linear structure of linear maps, we can obtain a different view of vectors as linear maps.

\begin{remark}[Linear operator view]
\label{rem-view}
For $\bu\in\cH$, let $\hat{\bu} : a \mapsto a\bu\in\cL(\bC;\cH)$, then its adjoint $\hat{\bu}^\dag\in\cL(\cH;\bC)$ has property $\hat{\bu}^\dag(\bv) = \<\bu,\bv\>$.
$\hat{\bu}$ is the operator view of $\bu$ and we do not distinguish between them (so $\bu$ refer to $\hat{\bu}$ when it is considered as linear map according to the context); $\hat{\bu}^\dag(\bv)$ is the dual of $\bu$, named as co-vectors and we simply write it as $\bu^\dag$ without ambiguity. Further, notice that inner and outer product can be understood as the composition: $[\bu,\bv] = \bu\circ\bv^\dag$ and $\<\bu,\bv\> = \bu^\dag\circ\bv$ since scalar $c\in\bC$ itself can be regarded as $\hat{c} : a\mapsto ca \in\cL(\bC)$.
\end{remark}

\section{Introduction to quantum mechanics}
Let us introduce quantum mechanics from a more concrete view with examples written in matrix form rather than abstract linear algebra; for more details, please refer to \cite{NC00}.
There are four postulates that relate linear algebra and quantum mechanics, provides the way to describe, understand, and analyse quantum systems and their evolution.

\begin{postulate}[State space, c.f. \cite{NC00}]
\label{postulate-state}
The state space of any isolated physical system is a complex vector space
with inner product (that is, a Hilbert space $\cH$). 
The system is completely described by its state vector, which is a unit vector in its state space.
\end{postulate}
For example, the state space of a qubit system is a two-dimensional Hilbert space $\cH_2$ with two orthonormal basis $|0\>=\matrix{1\\0}$ and $|1\>=\matrix{0\\1}$. Here, the Dirac notation $|\cdot\>$ represents vector, and we use $\<\cdot|$ to represent the costate (covector) -- conjugate-transpose of ket, e.g., $\<1| = |1\>^\dag=\matrix{0\\1}^\dag = \matrix{0 & 1}$. To model the probabilistic ensembles of quantum states, \emph{(partial) density matrices} are introduced, i.e., positive-semidefinite matrices on its state space with trace 1 (resp. $\le 1$). For example, suppose the system is in $|0\>$ or $|1\>$ with the same probability $0.5$, then the density matrix of the system is 
$$\rho=0.5|0\>\<0| + 0.5|1\>\<1| = 0.5\matrix{1\\0}\matrix{1 & 0}+ 0.5\matrix{0\\1}\matrix{0 & 1}=\matrix{0.5 & \\ & 0.5}.$$

\begin{postulate}[Evolution, c.f. \cite{NC00}]
\label{postulate-evolution}
The evolution of a closed quantum system is described by a unitary transformation\footnote{Such transformation is fully determined by Schr\"odinger equation of the system.}.
\end{postulate}
Suppose the evolution is a Hadamard gate $H = \frac{1}{\sqrt{2}}\matrix{1&1\\1&-1}$ and the initial state is $|0\>$, the final state is $H|0\> = \matrix{1&1\\1&-1}\matrix{1\\0} = \frac{1}{\sqrt{2}}\matrix{1\\1} = \frac{1}{\sqrt{2}}(|0\>+|1\>)$ which is commonly denoted by $|+\>$; in density matrix form, it described as 
$$H(|0\>\<0|)H^\dag = \frac{1}{\sqrt{2}}\matrix{1&1\\1&-1}\matrix{1 & \\ & 0} \left(\frac{1}{\sqrt{2}}\matrix{1&1\\1&-1}\right)^\dag = \frac{1}{2}\matrix{1 & 1\\ 1& 1} = |+\>\<+|.$$
More generally, the evolution of a open quantum system which might interact with the environment, is modelled as a quantum operation $\cE\in\cQO(\cH)$, i.e., a complete-positive trace-nonincreasing superoperator on $\cH$.

\begin{postulate}[Quantum measurement, c.f. \cite{NC00}]
\label{postulate-measure}
Quantum measurements are described by a collection $\{M_m\}$ of
measurement operators such that $\sum_mM_m^\dag M_m = I$, where $m$ refers to the possible measurement outcomes and $I$ the identity matrix. If the state is $|\psi\>$ immediately before the measurement then the probability that result $m$ occurs is $p(m) = \<\psi|M_m^\dag M_m|\psi\>$ and the post-measurement state is $\frac{M_m|\psi\>}{\sqrt{p(m)}}$. If the state is described by density operator $\rho$, then $p(m) = \tr(M_m^\dag M_m\rho)$ and the post-measurement state is $\frac{M_m\rho M_m^\dag}{\tr(M_m^\dag M_m\rho)}$.
\end{postulate}
An important example is the computational basis measurement, e.g., $M_b = \{M_0 = |0\>\<0|, M_1 = |1\>\<1|\}$ for qubit systems. For example, performing $M_b$ on state $|\psi\> = a|0\>+b|1\>$, the probability obtain outcome $0$ is $p(0) = \<\psi|M_m^\dag M_m|\psi\> = |a|^2$ with the post-measurement state being $\frac{a}{|a|}|0\>$.

\begin{postulate}[Composite systems, c.f. \cite{NC00}]
\label{postulate-composite}
The state space of a composite physical system is the tensor product of the state spaces of the component physical systems. Specifically, if system $i\in{1,2,\cdots,n}$ is prepared in $|\psi_i\>$, then the joint state of the total system is $\bigotimes_{i=1}^n|\psi_i\> = |\psi_1\>\otimes|\psi_2\>\otimes\cdots\otimes|\psi_n\>$.
\end{postulate}

According to Postulate \ref{postulate-state} (aka superposition principle), entangled states such as Bell state $|\Phi\> = \frac{1}{\sqrt{2}}(|0\>\otimes|0\>+|1\>\otimes|1\>)$ is a valid quantum state. We can do some simple calculation on $|\Phi\>$ to obtain some experience of entangled states. The matrix forms of $|\Phi\>$ and its density operator $|\Phi\>\<\Phi|$ are
\begin{align*}
  |\Phi\> & = \frac{1}{\sqrt{2}} \matrix{1\\0} \otimes \matrix{1\\0} + \matrix{0\\1} \otimes \matrix{0\\1}
  = \frac{1}{\sqrt{2}} \matrix{1\\0\\0\\0} + \matrix{0\\0\\0\\1} = \frac{1}{\sqrt{2}} \matrix{1\\0\\0\\1} \\
  |\Phi\>\<\Phi| & = \frac{1}{2} (|0\> \otimes |0\> + |1\> \otimes |1\>)(\<0| \otimes \<0| + \<1| \otimes \<1|) \\
                             & = \frac{1}{2} (|0\>\<0| \otimes |0\>\<0| + |0\>\<1| \otimes |0\>\<1| + |1\>\<0| \otimes |1\>\<0| + |1\>\<1| \otimes |1\>\<1|) \\
                             & = \frac{1}{2} \matrix{1 & 0 & 0 & 1\\ 0 & 0 & 0 & 0 \\ 0 & 0 & 0 & 0 \\ 1 & 0 & 0 & 1}.
\end{align*}
The reduced density operator of $|\Phi\>\<\Phi|$ on the first system is
\begin{align*}
  \rho_1 & = \tr_2(|\Phi\>\<\Phi|) \\
         & = \frac{1}{2} \big(|0\>\<0| \cdot \tr(|0\>\<0|) + |0\>\<1| \cdot \tr(|0\>\<1|) + |1\>\<0| \cdot \tr(|1\>\<0|) + |1\>\<1| \cdot \tr(|1\>\<1|) \big) \\
         & = \frac{1}{2} (|0\>\<0| + |1\>\<1|) \\
         & = \frac{1}{2} \matrix{1&0\\0&1}.
\end{align*}
This can be seen as the state of the first system when we discard the second.
Similarly, we can show that the reduced density operator on the second system is $\rho_2 = \rho_1$. However, their tensor product
\begin{align*}
  \rho_1 \otimes \rho_2 = \frac{1}{4} \matrix{1&0&0&0\\0&1&0&0\\0&0&1&0\\0&0&0&1} \neq |\Phi\>\<\Phi|
\end{align*}
This indicates that an entangled state in a composite system is not a simple composition of the each system.  
In fact, one can show that Bell state cannot be rewritten in the form of $|\Phi\>=|\phi_1\>\otimes|\phi_2\>$ or $|\Phi\>\<\Phi| = \rho_1 \otimes \rho_2$, i.e., it is not a joint of two independent states and this is what the word ``entangled'' refers. Entanglement is one of the essential phenomenon that disparity between classical and quantum physics. 

\section{Concrete syntax with data type}

In this section, we give more technical details about quantum variables and list other concrete syntax.
\begin{figure}
    \includegraphics[width=1\textwidth]{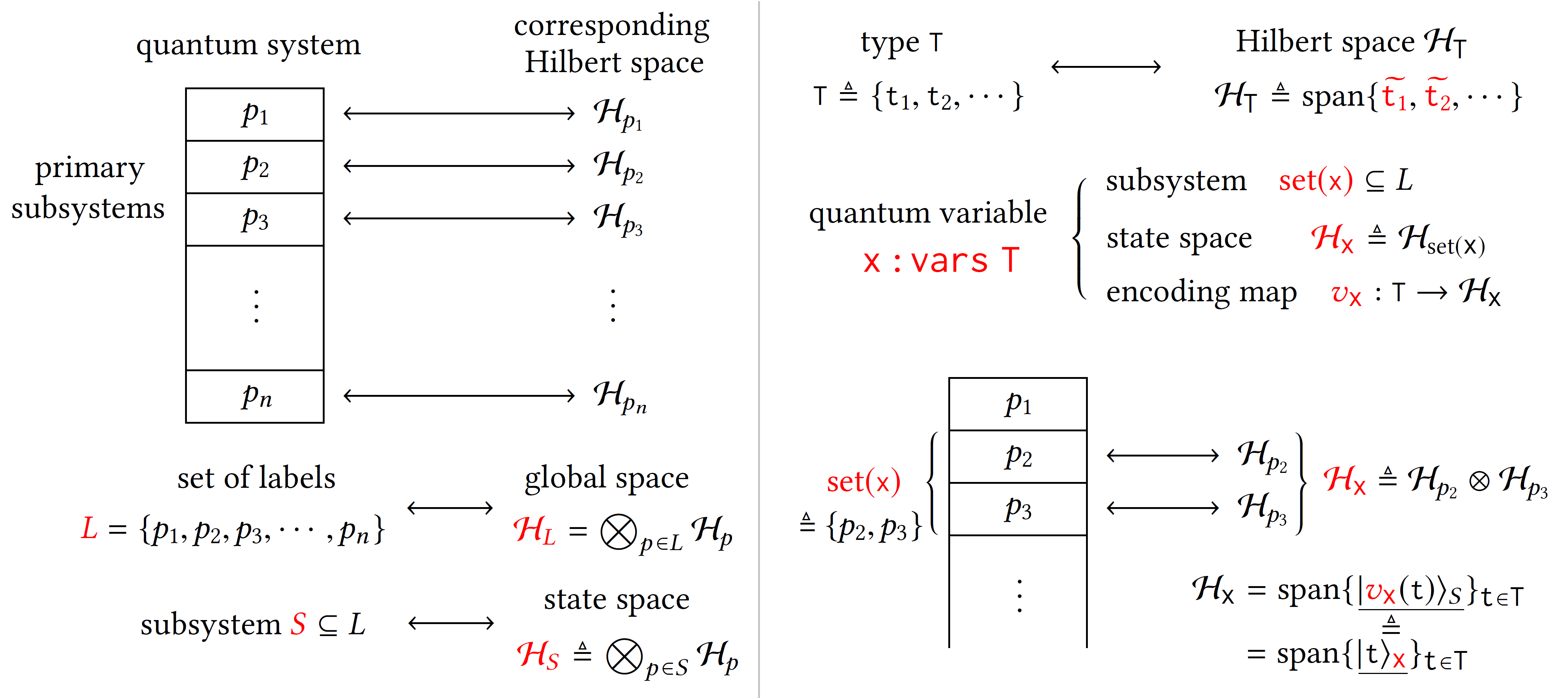}
    \caption{Global/local state space and mechanism of quantum variables.}
    \label{fig-variable}
\end{figure}

\subsection{Hilbert space associate with a data type} Let $\cH_\mT$ be a $|\mT|$-dimensional Hilbert space with the computational basis $\{\widetilde{\mt}\}_{\mt\in \mT}$ (i.e., $\<\widetilde{\mt_1},\widetilde{\mt_2}\> = 1$ if $\mt_1 = \mt_2$ and $0$ otherwise; we use tilde to denote that $\widetilde{\mt}\in\cH_\mT$ is a state), and we call $\cH_\mT$ the associated Hilbert space of $\mT$. 
Any state $\bu\in\cH_\mT$ can be written as 
$$\bu = \sum_{\mt\in \mT}\<\widetilde{\mt}, \bu\>\widetilde{\mt}$$
and linear operator $A$ on $\cH_\mT$ can be decomposed as 
$$A = \sum_{\mt_1,\mt_2\in \mT}\<\widetilde{\mt_2}, A(\widetilde{\mt_1})\>  [\widetilde{\mt_2},\widetilde{\mt_1}].$$

We further manipulate the type of tensor product.
Let $\mT_1\ast \mT_2$ be the product type of $\mT_1$ and $\mT_2$.
Realizing that $\cH_{\mT_1\ast\mT_2}$ is isomorphic to $\cH_{\mT_1}\otimes\cH_{\mT_2}$; we define the tensor product $\widetilde{\mt_1}\otimes\widetilde{\mt_2} \triangleq \widetilde{(\mt_1,\mt_2)} \in \cH_{\mT_1\ast\mT_2}$ and linearly extend to all states, i.e., for all $\bu_1\in\cH_{\mT_1}$ and $\bu_2\in\cH_{\mT_2}$, define:
$$\bu_1\otimes\bu_2 = \sum_{\mt_1,\mt_2\in \mT}\<\widetilde{\mt_1}, \bu_1\>\<\widetilde{\mt_2}, \bu_2\>\widetilde{(\mt_1,\mt_2)} \in \cH_{\mT_1\ast \mT_2}.$$
Similarly we can define the tensor product of operators $A_1\otimes A_2\in\cL(\cH_{\mT_1\ast\mT_2})$ by $A_1\otimes A_2 : \bu_1\otimes\bu_2 \mapsto A_1(\bu_1)\otimes A_2(\bu_2)$ for all $\bu_1\in\cH_{\mT_1}$ and $\bu_2\in\cH_{\mT_2}$ and then linearly extend to all states, that is,
$$A_1\otimes A_2 = \sum_{\mt_1,\ms_1\in \mT_1}\sum_{\mt_2,\ms_2\in \mT_s}\<\widetilde{\ms_1}, A_1(\widetilde{\mt_1})\>\<\widetilde{\ms_2}, A_2(\widetilde{\mt_2})\>  [\widetilde{(\ms_1,\ms_2)},\widetilde{(\mt_1,\mt_2)}].$$

Similarly, we can manipulate more complex dependent product type (or, $\Pi$ type).
CoqQ provide the tensor product dependent product type such as pair, tuple, finite function and dependent finite function.

\subsection{Quantum variables} 
A quantum variable of type $\mT$ associate its symbolic name and a storage location (quantum subsystem), together with the mapping between states in $\cH_\mT$ and states in its quantum subsystem. Formally, a quantum variable $\mx$ of type $\mT$, declared as \coqm{vars x : T}, consists of a subset $S\subseteq L$ denoted by $\pset{\mx}$ (we call it the domain of $\mx$), and an encoding function $v_\mx : \mT \rightarrow \cH_{\pset{\mx}}$ such that $\{|v_\mx(\mt)\>_\pset{\mx}\}_{\mt\in\mT}$ forms an orthonormal basis of $\cH_{\pset{\mx}}$. We can refer to Fig. \ref{fig-variable} for a picture. Intuitively, $\pset{\mx}$ indicates the subsystem that $\mx$ refers to, and $v_\mx$ tells us what is the computational basis of this system. We simply write $|\mt\>_\mx := |v_\mx(\mt)\>_\pset{\mx}$, which can be also realized as: inject $\widetilde{\mt}\in\cH_\mT$ to the subsystem $\pset{\mx}$ yields the state $|\mt\>_\mx\in\cH_\mx := \cH_\pset{\mx}$. More generally, we can inject any typed state or linear operator to $\mx$ as follows:
\begin{align*}
    \bu\in\cH_\mT \quad &\xrightarrow{\coqm{vars x : T}} \quad |u\>_\mx \triangleq \sum_{\mt\in \mT}\<\widetilde{\mt}, \bu\>|\mt\>_\mx\in\cH_\mx\\
    A\in\cL(\cH_\mT) \quad &\xrightarrow{\coqm{vars x : T}} \quad A[\mx] \triangleq \sum_{\mt_1,\mt_2\in \mT}\<\widetilde{\mt_2}, A(\widetilde{\mt_1})\>  |\mt_1\>_x\<\mt_2|\in\cL(\cH_\mx).
\end{align*}

\begin{figure}
    \centering
    \begin{align*}
    &\frac{\Gamma\vdash \coqm{x : vars T$_1$}\quad \Gamma\vdash \coqm{y : vars T$_2$}\quad \pset{\mx}\cap\pset{\my} = \emptyset}{\Gamma\vdash\coqm{[x,y] : vars (T$_1*$T$_2$)}}(\text{Var-Pair}) \\
    &\frac{
    \Gamma\vdash \coqm{x : n.-tuple (vars T)}\quad 
    \mi\neq \mj\rightarrow\pset{\mx(\mi)}\cap\pset{\mx(\mj)} = \emptyset
    }{
    \Gamma\vdash\coqm{$\overline{\mx}$ : vars (n.-tuple T)}
    }(\text{Var-tuple}) \\
    &\frac{
    \Gamma\vdash \coqm{x : F $\ra$ (vars T)}\quad 
    \mi\neq \mj\rightarrow\pset{\mx(\mi)}\cap\pset{\mx(\mj)} = \emptyset
    }{
    \Gamma\vdash\coqm{$\overline{\mx}$ : vars (F $\ra$ \mT)}
    }(\text{Var-FFun}) \\
    &\frac{
    \Gamma\vdash \coqm{x : $\Pi_{\mi :\, \coqm{F}}$(vars T$_\mi$)}\quad 
    \mi\neq \mj\rightarrow\pset{\mx(\mi)}\cap\pset{\mx(\mj)} = \emptyset
    }{
    \Gamma\vdash\coqm{$\overline{\mx}$ : vars $\Pi_{\mi :\, \coqm{F}}$\mT$_\mi$}
    }(\text{Var-Dep-FFun}) \\
    &\frac{\Gamma\vdash \coqm{x : vars T}\quad \Gamma\vdash \mt : \mT}{\Gamma\vdash \mx := |\mt\> : \coqm{cmd}}(\text{Init})\\
    &\frac{\Gamma\vdash \coqm{x : vars T}\quad \Gamma\vdash U : \cU(\cH_\mT)}{\Gamma\vdash \mx := U[\mx] : \coqm{cmd}}(\text{UT})\\
    &\frac{\Gamma\vdash \coqm{x : vars T$_1$}\quad \Gamma\vdash \coqm{x : vars T$_2$}\quad \Gamma\vdash U : \cU(\cH_{\mT_1\ast\mT_2})}{\Gamma\vdash [\mx,\my] := U[\mx,\my] : \coqm{cmd}}(\text{UT2})\\
    &\frac{\Gamma\vdash \coqm{x : vars T}\quad \Gamma\vdash C : \coqm{T $\rightarrow$ cmd}}{\Gamma\vdash \icond{\mt}{\rm meas}{\mx}{C} : \coqm{cmd}}(\text{Cond})\\
    &\frac{\Gamma\vdash \coqm{x : vars T}\quad \Gamma\vdash M : \coqm{QM(F;$\cH_\mT$)}\quad \Gamma\vdash C : \coqm{F $\rightarrow$ cmd}}{\Gamma\vdash \icond{\mt}{M}{\mx}{C} : \coqm{cmd}}(\text{CondG})\\
    &\frac{\Gamma\vdash \coqm{x : vars T}_1 \quad \Gamma\vdash\coqm{y : var T}_2\quad \Gamma\vdash C : \coqm{T}_1 * \coqm{T}_2 \rightarrow \coqm{cmd}}{\Gamma\vdash \icond{\mt}{\rm meas}{\mx,\my}{C} : \coqm{cmd}}(\text{Cond2})\\
    &\frac{\Gamma\vdash \coqm{x : vars T}_1 \quad \Gamma\vdash\coqm{y : var T}_2\quad \Gamma\vdash M : \coqm{QM(F;$\cH_{\mT_1\ast\mT_2}$)}\quad\Gamma\vdash C : \coqm{F $\rightarrow$ cmd}}{\Gamma\vdash \icond{\mt}{M}{\mx,\my}{C} : \coqm{cmd}}(\text{Cond2G})\\
    &\frac{\Gamma\vdash \coqm{x : vars }\mathbb{B} \quad \Gamma\vdash \coqm{b : }\mathbb{B}\quad \Gamma\vdash C : \coqm{cmd}}{\Gamma\vdash \iwhile{\rm meas}{\mx}{b}{C} : \coqm{cmd}}(\text{While})\\
    &\frac{\Gamma\vdash \coqm{x : vars T} \quad \Gamma\vdash M : \coqm{QM($\mathbb{B}$;$\cH_\mT$)}\quad\Gamma\vdash \coqm{b : }\mathbb{B}\quad \Gamma\vdash C : \coqm{cmd}}{\Gamma\vdash \iwhile{M}{\mx}{b}{C} : \coqm{cmd}}(\text{WhileG})\\
    &\frac{\Gamma\vdash \coqm{x : vars T}_1 \quad\Gamma\vdash \coqm{y : vars T}_2 \quad \Gamma\vdash M : \coqm{QM($\mathbb{B}$;$\cH_{\mT_1\ast\mT_2}$)}\quad\Gamma\vdash \coqm{b : }\mathbb{B}\quad \Gamma\vdash C : \coqm{cmd}}{\Gamma\vdash \iwhile{M}{\mx,\my}{b}{C} : \coqm{cmd}}(\text{While2G})
\end{align*}
    \caption{
    Typing rule of the concrete syntax, where
    \coqm{T}, \coqm{T}$_1$, \coqm{T}$_2$ are assumed to be inhabited finite type,  
    \coqm{F} is a finite type, \coqm{QM(F;}$\cH$\coqm{)} stands for the quantum measurement on $\cH$ with outcome set \coqm{F}.
    }
    \label{fig:typing_concrete_syntax}
\end{figure}

\subsection{Composition of quantum variables} Suppose two quantum variables \coqm{vars x$_1$ : T$_1$} and \coqm{vars x$_2$ : T$_2$} have disjoint domains, i.e., $\pset{\mx_1}\cap\pset{\mx_2} = \emptyset$. We manipulate the pair \coqm{[x$_1$,x$_2$]} as a quantum variable of type \coqm{T$_1\ast$T$_2$} with $\pset{[\mx_1,\mx_2]} = \pset{\mx_1}\cup\pset{\mx_2}$ by selecting:
$$v_{[\mx_1,\mx_2]} (\mt_1,\mt_2) := |\mt_1\>_{\mx_1}|\mt_2\>_{\mx_2} \in\cH_{\pset{\mx_1}\cup\pset{\mx_2}}; \quad \text{i.e.,}\quad |(\mt_1,\mt_2)\>_{[\mx_1,\mx_2]} = |\mt_1\>_{\mx_1}|\mt_2\>_{\mx_2}.$$
Such choice make the typed tensor product consistent to the composition of quantum variables; for example, suppose 
$|\phi_i\>\in\cH_{\mT_i}$ and $A_i\in\cL(\cH_{\mT_i})$ for $i=1,2$, we have
$$
|\phi_1\otimes\phi_2\>_{[\mx_1,\mx_2]} = |\phi_1\>_{\mx_1}|\phi_2\>_{\mx_2};\quad
(A_1\otimes A_2)[\mx_1,\mx_2] = A_1[\mx_1]\otimes A_2[\mx_2].
$$

\subsection{Concrete syntax}
Besides the concrete syntax that mentioned in main text, CoqQ provides more instantiation such as:
\begin{align*}
    & \mx := |\mt\>\ \big|\ \mx := U[\mx]\ \big|\ [\mx, \my] := U[\mx, \my] \\
    & |\ \icond{\mt}{M}{\mx}{C}\ \big|\ \icond{\mt}{M}{\mx,\my}{C} \\
    & |\ \icond{\mt}{\rm meas}{\mx}{C}\ \big|\ \icond{\mt}{\rm meas}{\mx,\my}{C} \\
    & |\ \iwhile{M}{\mx}{b}{C}\ \big|\ \iwhile{\mt}{M}{\mx,\my}{b}{C} \\
    & |\ \iwhile{\rm meas}{\mx}{b}{C}
\end{align*}
with (selected) typing rules in Fig. \ref{fig:typing_concrete_syntax}.

The concrete syntax is implemented as \emph{shallow embedding} in CoqQ; it is not limit on above syntax and typing rules. Furthermore, it reuses the type system of Coq since quantum variables share the same data type as in Coq; thus above typing rules are directly checked by Coq, i.e., we do not need to define the well-formedness of quantum programs.

\section{Inference rules}

Here, we list more selected rules proved in CoqQ as a supplementary to Fig. \ref{fig-rule}. As Fig. \ref{fig-rule-basic}, \ref{fig-rule-frame} and \ref{fig-rule-parallel} show, most of the rules are parameterized by ${\rm pt}$ and ${\rm st}$, saying these rules are sound for both partial and total correctness and saturated and non-saturated cases (non-saturated means may not be saturated).

\begin{figure}[h]\centering
\begin{equation*}\begin{split}
&({\rm Ax.Sk})\ \ \ \models_{\rm pt}^{\rm st}\{A\}\mathbf{Skip}\{A\} \qquad ({\rm Ax.In}) \ \ \ \models_{\rm pt}^{\rm st}\left\{{}_\mx\<\mt|A|\mt\>_\mx\right\}\mx := |\mt\>\{A\}\\
&({\rm Ax.InF}) \ \ \ \frac{S\cap\pset{x} = \emptyset}{\models_{\rm pt}^{\rm st}\{A_S\}\mx := |\mt\>\{A_S\otimes|\mt\>_\mx\<\mt|\}}\qquad
({\rm Ax.UT})\ \ \ \models_{\rm pt}^{\rm st}\{U[\mx]^{\dag}AU[\mx]\}\mx := U[\mx]\{A\}\\
&({\rm Ax.UTF})\ \ \ \models_{\rm pt}^{\rm st}\{A\}\mx := U[\mx]\{U[\mx]AU[\mx]^{\dag}\} \qquad 
({\rm R.SC})\ \ \ \frac{\models_{\rm pt}^{\rm st}\{A\}S_1\{B\}\quad\ \ \models_{\rm pt}^{\rm st}\{B\}S_2\{C\}}{\models_{\rm pt}^{\rm st}\{A\}S_1;S_2\{C\}}\\
&({\rm R.IF})\ \ \
\frac{\models_{\rm pt}^{\rm st}\{A_\mt\}C_\mt\{B\}\ {\rm for\ all}\ \mt}{\models_{\rm pt}^{\rm st}\big\{\sum_{\mt : \mT}
|\mt\>_\mx\<\mt|A_m|\mt\>_\mx\<\mt|\big\}\icond{\mt}{\rm meas}{\mx}{C}\{B\}}\\
&({\rm R.LP.P})\ \ \
\frac{R := |b\>_\mx\<b|A|b\>_\mx\<b|+|\neg b\>_\mx\<\neg b|B|\neg b\>_\mx\<\neg b|\quad \models_{\rm p}\{A\}C\{R\}\quad A\sqsubseteq I\quad B\sqsubseteq I}{\models_{\rm p}\{R\}
\iwhile{\rm meas}{\mx}{b}{C}\{B\}}\\
&({\rm R.Or})\ \ \ \frac{A\sqsubseteq
A^{\prime}\ \ \ \ \models_{\rm pt}\{A^{\prime}\}C\{B^{\prime}\}\ \ \ \
B^{\prime}\sqsubseteq B}{\models_{\rm pt}\{A\}C\{B\}}\\
&({\rm R.Scale.T})\ \ \ \frac{\models_{\rm t}^{\rm st}\{A\}C\{B\}\quad \lambda \ge 0}{\models_{\rm t}^{\rm st}\{\lambda A\}C\{\lambda B\}}\ \ \ \ 
({\rm R.Add.T})\ \ \ \frac{\models_{\rm t}^{\rm st}\{A_1\}C\{B_1\}\quad \models_{\rm t}^{\rm st}\{A_2\}C\{B_2\}}{\models_{\rm t}^{\rm st}\{A_1+A_2\}C\{B_1+B_2\}}\\
&({\rm R.CC.P})\ \ \ \frac{\forall i, \models_{\rm p}\{A_i\}C\{B_i\}\quad \forall i, 0\le\lambda_i\quad \sum_i\lambda_i\le 1}{\models_{\rm p}\{\sum_i\lambda_i A_i\}C\{\sum_i\lambda_i B_i\}}\\
&({\rm R.ST})\ \ \ \frac{\models_{\rm pt}^{\rm s}\{A\}C\{B\}}{\models_{\rm pt}\{A\}C\{B\}}\ \ \ \
({\rm R.No.LP})\ \ \ \frac{\models_{\rm pt}^{\rm st}\{A\}C\{B\}\quad C\text{\ has no while}}{\models_{\rm pt^\prime}^{\rm st}\{A\}C\{B\}} \\
&({\rm R.Lim})\ \ \ \frac{\lim_{n\rightarrow\infty} A_n = A\quad \forall n, \models_{\rm pt}^{\rm st}\{A_n\}C\{B_n\}\quad \lim_{n\rightarrow\infty}B_n = B}{\models_{\rm pt}^{\rm st}\{A\}C\{B\}}\\
&({\rm R.Inner})\ \ \ \frac{\models_{\rm t}^{\rm s}\{1\}C\{|v\>_{S_v}\<v|\}\quad \||v\>_{S_v}\|\le 1\quad S_u\subseteq S_v }{\models_{\rm t}^{\rm s}\{\|{}_{S_u}\<u|v\>_{S_v}\|^2\}C\{|u\>_{S_u}\<u|\}}\\
&({\rm R.Inner})\ \ \ \frac{\models_{\rm t}^{\rm s}\{1\}C\{|v\>_S\<v|\}\quad \||v\>_S\|\le 1 }{\models_{\rm t}^{\rm s}\{\|\<u|v\>\|^2\}C\{|u\>_S\<u|\}}\\
\end{split}\end{equation*}
\caption{Selected inference rules for basic construct and structural rules. $\rm{pt} \in \{{\rm p}, {\rm t}\}$ indicates the total/partial correctness; $\rm{st} \in \{{\rm s},{\rm n}\}$ to indicates the correctness formula is saturated or not (we write $\models$ for $\models^{\rm n}$).}
\label{fig-rule-basic}
\end{figure}

\begin{figure}[h]\centering
\begin{equation*}\begin{split}
&({\rm Ax.Inv})\ \ \ \frac{A_S\sqsubseteq I_S\quad S\cap \pset{C} = \emptyset}{\models_{\rm p}\{A_S\}C\{A_S\}}\qquad ({\rm R.SO})\ \ \ \frac{\models_{\rm pt}^{\rm st}\{A\}C\{B\}\quad S\cap \pset{C} = \emptyset\quad \cE_S\in\cQC(\cH_S)}{\models_{\rm pt}^{\rm st}\{\cE_S^\ast(A)\}C\{\cE_S^\ast(B)\}}\\
&({\rm R.El})\ \ \ \frac{\models_{\rm pt}^{\rm st}\{A_{S_A}\}C\{B\}\quad S_A\cap S = \emptyset}{\models_{\rm pt}^{\rm st}\{A_{S_A}\otimes I_{S}\}C\{B\}}\ \ \ \ 
({\rm R.Er})\ \ \ \frac{\models_{\rm pt}^{\rm st}\{A\}C\{B_{S_B}\}\quad S\cap S_B = \emptyset}{\models_{\rm pt}^{\rm st}\{A\}C\{B_{S_B}\otimes I_{S}\}}\\
&({\rm R.TI})\ \ \ \frac{\models_{\rm pt}^{\rm st}\{A_{S_A}\otimes I_{S_1}\}C\{B_{S_B}\otimes I_{S_2}\}\quad S_A\cap S_1=\emptyset\quad S_B\cap S_2=\emptyset}{\models_{\rm pt}^{\rm st}\{A_{S_A}\}C\{B_{S_B}\}}\\
&({\rm Frame.T})\ \ \ \frac{\models_{\rm t}^{\rm st}\{A_{S_A}\}C\{B_{S_B}\}\quad 0\sqsubseteq R_S\quad (\pset{C}\cup S_A\cup S_B)\cap S = \emptyset\quad}{\models_{\rm t}^{\rm st}\{A_{S_A}\otimes R_{S}\}C\{B_{S_B}\otimes R_{S}\}}\\
&({\rm Frame.P})\ \ \ \frac{\models_{\rm p}\{A_{S_A}\}C\{B_{S_B}\}\quad 0\sqsubseteq R_S\sqsubseteq I_S\quad (\pset{C}\cup S_A\cup S_B)\cap S = \emptyset\quad}{\models_{\rm p}\{A_{S_A}\otimes R_{S}\}C\{B_{S_B}\otimes R_{S}\}}
\end{split}\end{equation*}
\caption{Selected inference rules for manipulating local reasoning. $\cQC$ is the set of quantum channels on $\cH_S$, i.e., the set of completely-positive and trace-preserving super-operators.}
\label{fig-rule-frame}
\end{figure}

\begin{figure}[h]\centering
\begin{equation*}\begin{split}
&({\rm R.PC.T})\frac{\begin{split}
\forall i, \models_{\rm t}^{\rm st}\{A_{i,S_{A_i}}\}P_i\{B_{i,S_{B_i}}\}\quad \forall i, 0\sqsubseteq A_{i,S_{A_i}} \quad \forall i, 0\sqsubseteq B_{i,S_{B_i}}\ \ \\
\forall i\neq j, (\pset{C_i}\cup S_{A_i}\cup S_{B_i}) \cap (\pset{C_j}\cup S_{A_j}\cup S_{B_j}) = \emptyset
	\end{split}}{\models_{\rm t}^{\rm st}\{\bigotimes_iA_{i,S_{A_i}}\}\mathbf{for}\ i\ \mathbf{do}\ C_i\{\bigotimes_iB_{i,S_{B_i}}\}}\\
&({\rm R.PC.P})\frac{\begin{split}
		\forall i, \models_{\rm p}\{A_{i,S_{A_i}}\}P_i\{B_{i,S_{B_i}}\}\quad \forall i, 0\sqsubseteq A_{i,S_{A_i}}\sqsubseteq I_{S_{A_i}} \quad \forall i, 0\sqsubseteq B_{i,S_{B_i}}\sqsubseteq I_{S_{B_i}}\\
		\forall i\neq j, (\pset{C_i}\cup S_{A_i}\cup S_{B_i}) \cap (\pset{C_j}\cup S_{A_j}\cup S_{B_j}) = \emptyset\qquad\ \ 
\end{split}}{\models_{\rm p}\{\bigotimes_iA_{i,S_{A_i}}\}\mathbf{for}\ i\ \mathbf{do}\ C_i\{\bigotimes_iB_{i,S_{B_i}}\}}\\
&({\rm Ax.UTP})\ \ \ \frac{\forall i\neq j, \pset{\mx_i} \cap \pset{\mx_j} = \emptyset}{\models_{\rm pt}^{\rm st}\{(\bigotimes_iU_i[\mx_i])^\dag A (\bigotimes_iU_i[\mx_i])\}\mathbf{for}\ i\ \mathbf{do}\ \mx_i := U_i[\mx_i]\{A\}}\\
&({\rm Ax.UTFP})\ \ \ \frac{\forall i\neq j, \pset{\mx_i} \cap \pset{\mx_j} = \emptyset}{\models_{\rm pt}^{\rm st}\{A\}\mathbf{for}\ i\ \mathbf{do}\ \mx_i := U_i[\mx_i]\{(\bigotimes_iU_i[\mx_i])A(\bigotimes_iU_i[\mx_i])^\dag\}}\\
&({\rm Ax.InP})\ \ \ \frac{\forall i\neq j, \pset{\mx_i} \cap \pset{\mx_j} = \emptyset}{\models_{\rm pt}^{\rm st}\{\prod_i\<\mt_i|A_i|\mt_i\>\}\mathbf{for}\ i\ \mathbf{do}\ \mx_i := |\mt_i\>\{\bigotimes_iA_i[\mx_i]\}}\\
&({\rm Ax.InFP})\ \ \ \frac{\forall i\neq j, \pset{\mx_i} \cap \pset{\mx_j} = \emptyset}{\models_{\rm pt}^{\rm st}\{1\}\mathbf{for}\ i\ \mathbf{do}\ \mx_i := |\mt_i\>\{\bigotimes_i|\mt_i\>_{\mx_i}\<\mt_i|\}}
\end{split}\end{equation*}
\caption{Selected inference rules to facilitate parallel reasoning for \textbf{for} loops.}
\label{fig-rule-parallel}
\end{figure}

\end{appendix}

\end{document}